\documentclass[prx,twocolumn,superscriptaddress,citeautoscript,showpacs,amsart,longbibliography]{revtex4-1}

\usepackage{graphicx}
\usepackage{multirow}
\usepackage{color}
\usepackage{bm}
\usepackage{times}
\usepackage{amsmath,bm,amsfonts}
\usepackage{dcolumn}
\usepackage{graphicx}
\usepackage{latexsym}

\usepackage{ulem} 
\usepackage{braket}

\usepackage{cancel}

\def\Pf{{\rm Pf}}
\def\d{\partial}

\begin{document}
\title{
Failure of Nielsen-Ninomiya theorem and fragile topology in two-dimensional systems with space-time inversion symmetry: Application to twisted bilayer graphene at magic angle
}

\author{Junyeong \surname{Ahn}}
\thanks{These authors contributed equally to this work.}
\affiliation{Department of Physics and Astronomy, Seoul National University, Seoul 08826, Korea}

\affiliation{Center for Correlated Electron Systems, Institute for Basic Science (IBS), Seoul 08826, Korea}

\affiliation{Center for Theoretical Physics (CTP), Seoul National University, Seoul 08826, Korea}

\author{Sungjoon \surname{Park}}
\thanks{These authors contributed equally to this work.}
\affiliation{Department of Physics and Astronomy, Seoul National University, Seoul 08826, Korea}

\affiliation{Center for Correlated Electron Systems, Institute for Basic Science (IBS), Seoul 08826, Korea}

\affiliation{Center for Theoretical Physics (CTP), Seoul National University, Seoul 08826, Korea}

\author{Bohm-Jung \surname{Yang}}
\email{bjyang@snu.ac.kr}
\affiliation{Department of Physics and Astronomy, Seoul National University, Seoul 08826, Korea}

\affiliation{Center for Correlated Electron Systems, Institute for Basic Science (IBS), Seoul 08826, Korea}

\affiliation{Center for Theoretical Physics (CTP), Seoul National University, Seoul 08826, Korea}

\date{\today}

\begin{abstract}
We show that the Wannier obstruction and the fragile topology of the nearly flat bands in twisted bilayer graphene at magic angle are manifestations of the nontrivial topology of two-dimensional real wave functions characterized by the Euler class. To prove this, we examine the generic band topology of two dimensional real fermions in systems with space-time inversion $I_{ST}$ symmetry. The Euler class is an integer topological invariant classifying real two band systems. We show that a two-band system with a nonzero Euler class cannot have an $I_{ST}$-symmetric Wannier representation. Moreover, a two-band system with the Euler class $e_{2}$ has band crossing points whose total winding number is equal to $-2e_2$. Thus the conventional Nielsen-Ninomiya theorem fails in systems with a nonzero Euler class. We propose that the topological phase transition between two insulators carrying distinct Euler classes can be described in terms of the pair creation and annihilation of vortices accompanied by winding number changes across Dirac strings. When the number of bands is bigger than two, there is a $Z_{2}$ topological invariant classifying the band topology, that is, the second Stiefel Whitney class ($w_2$). Two bands with an even (odd) Euler class turn into a system with $w_2=0$ ($w_2=1$) when additional trivial bands are added. Although the nontrivial second Stiefel-Whitney class remains robust against adding trivial bands, it does not impose a Wannier obstruction when the number of bands is bigger than two. However, when the resulting multi-band system with the nontrivial second Stiefel-Whitney class is supplemented by additional chiral symmetry, a nontrivial second-order topology and the associated corner charges are guaranteed.
\end{abstract}

\maketitle


\section{Introduction}

The recent discovery of Mott insulating states and superconductivity in twisted bilayer graphene (TBG) near the first magic angle $\theta\sim1.05^{\circ}$~\cite{cao2018unconventional, cao2018correlated} has lead to a surge of research activities to understand this system~\cite{volovik2018graphite,xu2018topological,roy2018unconventional,guo2018pairing,baskaran2018theory,padhi2018wigner,irkhin2018dirac,dodaro2018phases,huang2018antiferromagnetically,zhang2018low,ray2018wannier, liu2018chiral,xu2018kekul,rademaker2018charge,isobe2018superconductivity, wu2018theory,pizarro2018nature,peltonen2018mean,you2018superconductivity,choi2018strong,wu2018emergent, pal2018magic,ochi2018possible,fidrysiak2018unconventional,thomson2018triangular, guinea2018electrostatic,zou2018band,po2018origin,koshino2018maximally,yuan2018model,kang2018symmetry}. One notable feature in the band structure of TBG is the presence of almost flat bands near charge neutrality, which are effectively decoupled from other bands by an energy gap. The reduced kinetic energy of the flat bands allows this purely carbon-based system, normally regarded as a weakly correlated system, to be an intriguing playground to examine the Mott physics and the associated unconventional superconductivity. 

For a microscopic description of correlation effect in TBG, there have been several theoretical efforts to construct a tight-binding lattice model capturing the characteristic band structure of the four almost flat bands near charge neutrality~\cite{zou2018band,yuan2018model,po2018origin,kang2018symmetry,koshino2018maximally}. Here we neglect the spin degrees of freedom for counting the number of bands, which is valid because the spin-orbit coupling is negligibly small. According to the low energy continuum theory which excellently describes the qualitative feature of the almost flat bands, there are two Dirac points at each K and K' point in the Moir\'e Brillouin zone, whose origin can be traced back to the Dirac points at each valley of the underlying graphene layers~\cite{dos2007graphene,bistritzer2011moire}. The presence of massless Dirac fermions is further supported by several theoretical studies~\cite{shallcross2010electronic,suarezmorell2010flat,trambly2010localization,jung2014ab} as well as recent quantum oscillation measurement~\cite{cao2016superlattice}.
The existence of gapless Dirac points indicates the $U_{v}(1)$ valley and the space-time inversion $C_{2z}T$ symmetries, where $C_{2z}$ denotes a two-fold rotation about the $z$-axis and $T$ is time-reversal symmetry~\cite{po2018origin,zou2018band}. 
In the presence of $U_{v}(1)$ and $C_{2z}T$ symmetries, the four nearly flat bands are decoupled into two independent valley-filtered two-band systems, and each two-band system possesses Dirac points at $K$ and $K'$.
The fact that both the valley charge conservation and $C_{2z}T$ symmetries are not the exact symmetry of the TBG indicates that the symmetry of the low energy physics is larger than the exact lattice symmetry~\cite{zou2018band}.

Interestingly, by putting together all the emergent symmetries including $U_{v}(1)$ and $C_{2z}T$ symmetries, Po et al. have found an obstruction to constructing well-localized Wannier functions describing the four nearly flat bands in TBG~\cite{po2018origin,zou2018band}.
Moreover, it has been shown that the obstruction originates from the fact that the two Dirac points in each valley-filtered two-band system have the same winding number, which is generally not allowed in 2D periodic systems due to the Nielsen-Ninomiya theorem~\cite{nielsen1981no}. In addition, based on the observation that the winding number is defined only for a two-band model in each valley, it is conjectured that the Wannier obstruction is fragile~\cite{po2018origin,zou2018band}, that is, the obstruction disappears after one or more trivial (i.e., Wannier-representable) bands are added to the model.

The main purpose of the present study is to unveil the topological nature of the nearly flat bands in TBG near a magic angle and propose a general framework to understand the band topology of 2D systems sharing the same symmetry. In particular, we show that, two bands having two Dirac points with the same winding number are endowed with an integer topological invariant, the Euler class $e_2$, when a 2D spinless fermion system has symetry under space-time inversion $I_{ST}\equiv C_{2z}T$. We explicitly show that two bands having a nonzero Euler class cannot have an exponentially localized Wannier representation, that is, there is a Wannier obstruction. Moreover, the nonzero Euler class $e_2$ implies that there are band crossing points, henceforth called vortices, between the two bands, whose total winding number is equal to $-2e_2$. Thus, a real two-band system carrying a nonzero $e_2$ evidences the violation of the Nielsen-Ninomiya theorem.

When the number of occupied bands is bigger than two, the system is characterized by another $Z_{2}$ topological invariant, that is, the second Stiefel-Whitney (SW) class $w_{2}$. A two-band system with the Euler class $e_2$ turns into a multi-band system with the second Stiefel-Whitney class $w_2=e_2$ (mod 2) when additional trivial bands are added. Therefore a two band system with an odd $e_2$ can still be characterized by the nontrivial $w_2=1$. Even if $w_2$ is nontrivial in a multi-band system, we show that there is no obstruction to constructing well-localized Wanner functions, which indicates that the band topology of a real two band system is fragile.
When a multi-band system with $w_2=1$ has additional chiral symmetry, the system exhibits a higher-order band topology which guarantees the existence of corner charges.

This paper is organized as follows. We first present the topological properties
of a simple four-band lattice model proposed by Zou et al.~\cite{zou2018band}, which captures all the essential properties of the nearly flat bands in TBG at magic angle. After clarifying
the issues related with the band topology of TBG by using the toy model, we provide a general description of the band topology of space-time-inversion-symmetric spinless fermion systems in 2D.
We review the concept of the Euler class in Sec.~\ref{sec.EulerSW} and show that a nontrivial Euler class leads to a Wannier obstruction in Sec.~\ref{sec.Wannier}.
In Sec~\ref{sec.Nielsen}. we prove the correspondence between the vortex winding number and the Euler class, and demonstrate the violation of Nielsen-Ninomiya theorem in real two band systems with a nonzero Euler class.
Based on this correspondence, we study the transition changing the Euler class by analyzing the winding number of vortices.
In Sec.~\ref{sec.off-diagonal}, we develop a new method for calculating the winding number by using the off-diagonal component of the Berry connection. We use this to study the pair creation and annihilation of vortices in Sec.~\ref{sec.nonorientable}, where we show that the topological phase transition between two insulators carrying distinct Euler classes can be described via the pair creation and annihilation of vortices through the winding number reversal across a Dirac string.
In Sec.~\ref{sec.fragile}, we describe the fragile and higher-order nature of the band topology of the nearly flat bands in twisted bilayer graphene based on the winding number annihilation and the properties of the second Stiefel-Whitney class.
In Sec.IX, we summarize the main results and discuss future research directions.
In addition, we explain how the winding number can be computed by using the off-diagonal Berry phase in a generic chiral symmetric system in Appendix \ref{sec.chiral}. In Appendix.~\ref{sec.monopole}, we discuss the vortex annihilation in the point of view of monopole nodal lines. Appendix \ref{sec.tpt_soc}, we show the equivalence between the second Stiefel-Whitney class and the Fu-Kane-Mele invariant in spin-orbit coupled two dimensional systems with $I_{ST}=C_{2z}T$ symmetry. Finally, in Appendix \ref{sec.SW-SOTI}, we explain the symmetry protection of anomalous corner states and review the Wilson loop method to characterize the second-order band topology by extracting the first and second Stiefel Whitney classes directly from the Wilson loop spectrum without additional numerical computation of the nested Wilson loop.

\section{Band topology of nearly flat bands in twisted bilayer graphene}

Let us first clarify the issues related with the band topology of the nearly flat bands in TBG at magic angle. For this purpose, we study a simple four-band model Hamiltonian proposed by Zou et al.~ \cite{zou2018band}, which captures the essential characteristics of the nearly flat bands in TBG.

\begin{figure}[t!]
\includegraphics[width=8.5cm]{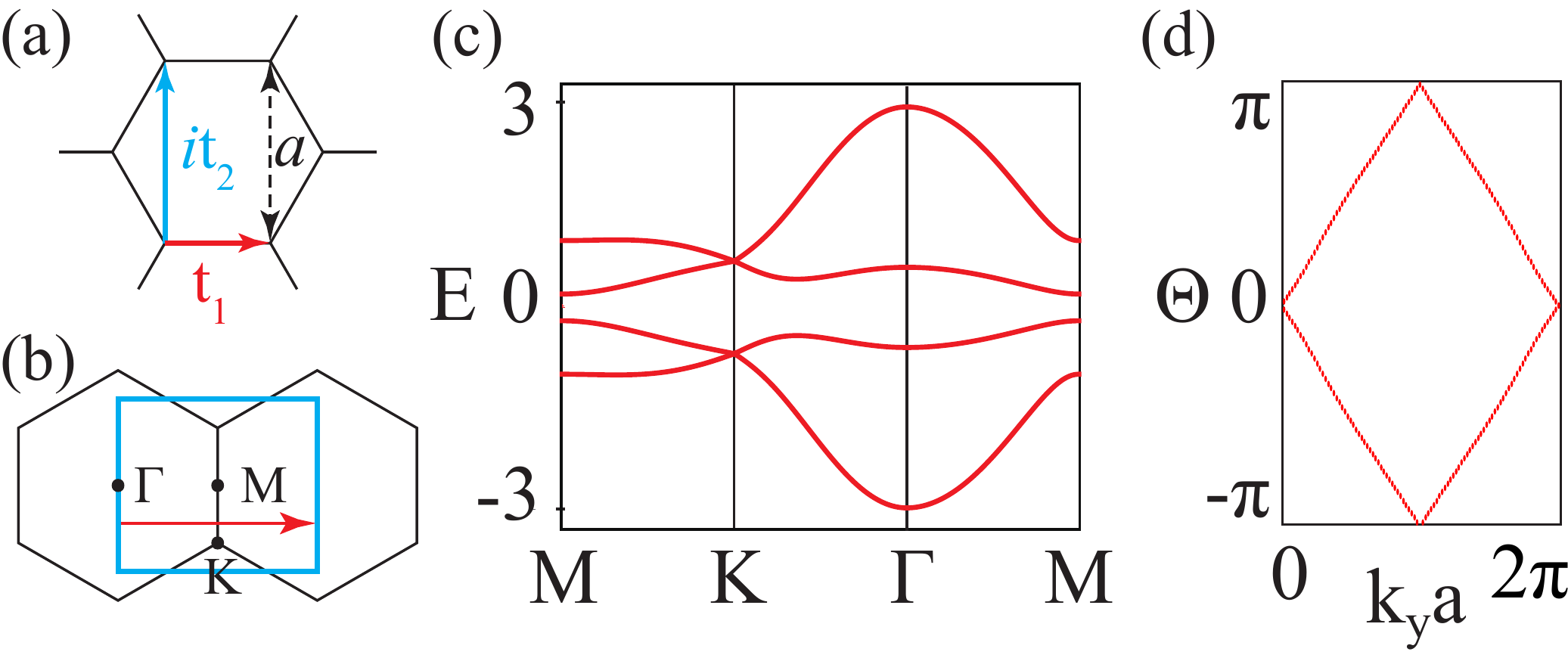}
\caption{
(a) The definition of the hopping amplitudes $t_{1,2}$ for a four-band lattice model proposed by Zou et. al.~\cite{zou2018band}. $a$ is the lattice constant representing the
lattice vector for a Moir\'e superlattice of TBG at magic angle.
(b) High symmetry points in the Brillouin zone.
$\Gamma=(0,0)$, $M=(2\pi/\sqrt{3}a,0)$, and $K=(2\pi/\sqrt{3}a,2\pi/3a)$.
The blue rectangle is the Brillouin zone used to compute the Wilson loop spectrum.
(c) Band structure along high-symmetry lines.
Both of the occupied and unoccupied bands have gapless Dirac points at K and K'.
(d) Wilson loop spectrum for the lower two bands.
The Wilson loop operator is calculated along the $k_y$ direction at fixed $k_x$, as shown by the red arrow in (b).
The unit winding of the spectrum indicates the unit Euler class $|e_2|=1$.
}
\label{fig.TB}
\end{figure}

\subsection{A four-band lattice model}

The model is defined on a honeycomb lattice which represents the Moir\'e superlattice of TBG at magic angle~\cite{zou2018band}. Putting two orbitals per site, one can construct a four-band Hamiltonian given by 
\begin{align}
\label{eq.TB}
H
=\sum_{\braket{ij}}c^{\dagger}_i(\hat{t}_1)_{ij}c_j+\sum_{\braket{\braket{ij}}}c^{\dagger}_is_{ij}(i\hat{t}_2)_{ij}c_j,
\end{align}
where $\hat{t}_1=0.4+0.6\tau_z$ and $\hat{t}_2=0.1\tau_x$ indicate the hopping amplitudes between the nearest-neighbor and next-nearest neighbor sites with the Pauli matrices $\tau_{x,y,z}$ representing the orbital degrees of freedom. Here we choose $s_{ij}=+1$ for ${\bf r}_i={\bf r}_j+a\hat{y}$, which determines the rest of the $s_{ij}$'s because of the $C_{3z}$ symmetry. Then the full Hamiltonian is invariant under a three-fold rotation about the $z$-axis $C_{3z}$, a two-fold rotation about the $y$-axis $C_{2y}$, and $C_{2z}T$. Namely, the lattice model has $D_{6}$ point group symmetry. 
This model Hamiltonian inherits the essential features of the nearly flat bands of TBG with enlarged emergent symmetries.
In momentum space, the Hamiltonian becomes
\begin{align}
H({\bf k})
&=\hat{t}_1\Bigg[\left(1+2\cos\frac{\sqrt{3}k_xa}{2}\cos\frac{k_ya}{2}\right)\sigma_x\notag\\
&\qquad\quad\;\,+2\sin\frac{\sqrt{3}k_xa}{2}\cos\frac{k_ya}{2}\sigma_y\Bigg]\notag\\
&+\hat{t}_2\left(4\cos\frac{\sqrt{3}k_xa}{2}\sin\frac{k_ya}{2}-2\sin k_ya\right),
\end{align}
where the Pauli matrices $\sigma_{x,y,z}$ denote the sublattice degrees of freedom of the honeycomb lattice.

\subsection{Band topology of lower two bands}

The band structure of the four-band model is shown in Fig.~\ref{fig.TB}(c).
One can see that two lower bands are fully separated from the two upper bands. 
The two lower bands cross at two corners of the BZ, $K$ and $K'$, forming two Dirac points with the same winding number. As pointed out in \cite{zou2018band}, the winding number of the two Dirac points can be determined by examining the mirror eigenvalues of the two occupied bands at the $M$ point: if their mirror eigenvalues are opposite (equal), the winding numbers of the Dirac points at $K$ and $K'$ points are equal (opposite).  In the case of the model Hamiltonian in Eq.~(\ref{eq.TB}), the mirror symmetry along the $\Gamma M$ line can be represented by $\tau_z$, and it can be explicitly checked that the mirror eigenvalues of the two occupied bands are indeed opposite along this line. Similarly, the two upper bands also possess two Dirac points sharing the same winding number whose winding direction is opposite to that between the lower two bands. Both the lower two bands and the upper two bands possess the same topological characteristic of the nearly flat bands of TBG in a single valley while preserving all $D_{6}$ point group symmetry~\cite{zou2018band}.

Let us focus on the topological properties of the lower two bands to understand the band topology and the relevant obstruction of the nearly flat bands in TBG. 
One direct evidence showing the nontrivial topology of the lower two bands is the winding of the Wilson loop spectrum shown in Fig.~\ref{fig.TB}(d), which is computed from the transition function in a real gauge by using the technique developed in~\cite{ahn2018linking}.
Here the Wilson loop operator corresponds to the transition function.
In the Wilson loop spectrum in Fig.~\ref{fig.TB}(d), two eigenvalues change symmetrically about $\Theta=0$ line due to the $I_{ST}$ symmetry, and each eigenvalue winds once as $k_{y}$ is varied.
Below we show that the unit winding of the transition function in a real gauge indicates the unit Euler class $|e_2|=1$, which imposes an obstruction to Wannier representation and leads to the violation of the Nielsen-Ninomiya theorem.

\section{Euler class and Wannier obstruction for real fermions in two dimensions}
\label{sec.EulerSW}

The central symmetry governing the band topology of nearly flat bands in TBG is the symmetry under space-time inversion $I_{ST}$. $I_{ST}$ is an antiunitary symmetry operator, local in momentum space satisfying $I_{ST}^{2}=+1$, so it acts like a complex conjugation in momentum space~\cite{fang2015new}. In the absence of spin-orbit coupling, either $PT$ or $C_{2z}T$ can be used to define $I_{ST}$ where $P$ indicates a spatial inversion, $C_{2z}$ is a two-fold rotation about the $z$-axis, and $T$ is time reversal.
On the other hand, in the presence of spin-orbit coupling, only $C_{2z}T$ can be used to define $I_{ST}$ since $(PT)^{2}=-1$~\cite{fang2015new,ahn2017unconventional,kim2017two}.
In an $I_{ST}$ invariant system, we define a real gauge as
\begin{align}
\label{eq.reality}
I_{ST}\ket{\tilde{\psi}_{n\bf k}}=\ket{\tilde{\psi}_{n\bf k}},
\end{align}
where $\ket{\tilde{\psi}_{n\bf k}}$ is a Bloch state.
Other possible choices of real gauges are related to each other by orthogonal transformations.
This gauge condition is equivalent to $I_{ST}\ket{\tilde{u}_{n\bf k}}=\ket{\tilde{u}_{n\bf k}}$ for the cell-periodic part $\ket{\tilde{u}_{n\bf k}}= e^{-i{\bf k}\cdot \hat{\bf r}}\ket{\tilde{\psi}_{n\bf k}}$ since $e^{-i{\bf k}\cdot \hat{\bf r}}$ commutes with $I_{ST}$.
Moreover, the transition function of $\ket{\tilde{\psi}_{n\bf k}}$ is equivalent to that of $\ket{\tilde{u}_{n\bf k}}$ if we define the periodic condition to be $\ket{\tilde{\psi}_{n\bf k+G}}=\ket{\tilde{\psi}_{n\bf k}}$ and $\ket{\tilde{u}_{n\bf k+G}}=e^{i{\bf G}\cdot r}\ket{\tilde{u}_{n\bf k}}$, respectively.
As is customary, we will investigate the topology of the Bloch states using their cell-periodic part.
In this section, we define a topological invariant of $I_{ST}$-symmetric two-band systems, that is, the Euler class, and explain the topological obstruction for real states arising from it.

\subsection{The Euler class}

The Euler class $e_2$ is an integer topological invariant for two real bands which can be written as a simple flux integral form~\cite{hatcher2003vector,zhao2017p,nakahara2003geometry},
\begin{align}
\label{eq.Euler}
e_2=\frac{1}{2\pi}\oint_{BZ}d{\bf S}\cdot\tilde{\bf F}_{12},
\end{align}
where $
\tilde{\bf F}_{mn}({\bf k})
=\nabla_{\bf k}\times \tilde{\bf A}_{mn}({\bf k})
$
and
$
\tilde{\bf A}_{mn}({\bf k})
=\braket{\tilde{u}_m({\bf k})|\nabla_{\bf k}|\tilde{u}_n({\bf k})}
$
($m,n=1,2$) are $2\times 2$ antisymmetric real Berry curvature and connection defined by real states $\ket{\tilde{u}_n({\bf k})}$ in Eq.~(\ref{eq.reality}).
It is invariant under any $SO(2)$ gauge transformation, which has the form $O({\bf k})=\exp [-i\sigma_y\phi({\bf k})]$ and satisfies $\det[O({\bf k})]=1$.
On the other hand, under an orientation-reversing transformation with $\det[O({\bf k})]=-1$, which has the form $O({\bf k})=\sigma_z\exp [-i\sigma_y\phi({\bf k})]$,
$e_{2}$ changes its sign.
Therefore, the Euler class is well-defined only for orientable real states, that is, the states associated only with $O({\bf k})$ with the unit determinant.

The flux integral form of $e_2$ can be connected to transition functions in the following way.
To show this relation, let us notice that the 2D Brillouin zone can be deformed to a sphere when the real states are orientable along any non-contractible one-dimensional cycles as far as the topology of the real states is concerned [See Fig.~\ref{fig.patch}]. Then the sphere can be divided into two hemispheres, the northern ($N$) and southern ($S$) hemispheres, which overlap along the equator. Along the equator, the real smooth wave functions 
$\ket{u^{N}}$ and $\ket{u^{S}}$ defined on the northern and southern hemispheres, respectively can be connected 
by a transition function $t^{NS}=\braket{u^N|u^S}=\exp [-i\sigma_y\phi_{NS}]\in SO(2)$.
It is straightforward to show that
\begin{align}\label{eqn:e2_transition}
e_2
&=\frac{1}{2\pi}\oint_{S^2} d{\bf S}\cdot \tilde{\bf F}_{12}\notag\\
&=\frac{1}{2\pi}\int_{N} d{\bf S}\cdot \tilde{\bf F}_{12}+\frac{1}{2\pi}\int_{S} d{\bf S}\cdot \tilde{\bf F}_{12}\notag\\
&=\frac{1}{2\pi}\oint_{S^1}d{\bf k}\cdot \left(\tilde{\bf A}_{N,12}- \tilde{\bf A}_{S,12}\right)\notag\\
&=\frac{1}{2\pi}\oint_{S^1}d{\bf k}\cdot \nabla_{\bf k}\phi_{NS},
\end{align}
where $S^1$ indicates the circle along the equator.
Therefore the Euler class $e_{2}$ is identical to the winding number of the transition function $t^{NS}$. 

Let us note that Eq.~(\ref{eqn:e2_transition}) is also equivalent to the definition of the monopole charge~\cite{fang2015topological,zhao2017p}

\subsection{Wannier obstruction from the Euler class} 
\label{sec.Wannier}

Here we show that two real bands with a nontrivial Euler class suffer from an obstruction to defining exponentially localized Wannier functions respecting $I_{ST}$ symmetry.
Below, we prove the contrapositive, that the existence of exponentially localized $I_{ST}$-symmetric Wannier functions implies that the Euler class is trivial. Our strategy for the proof is to start from the $I_{ST}$-symmetric exponentially localized Wannier representation. Then, we go to the Bloch representation, find the transformation that makes $I_{ST}=K$, and finally determine whether a transition function with a nonzero winding number can arise in this real basis.

Let us recall some basic facts.
Wannier states $\ket{w_{n{\bf R}}}$ are defined to be the Fourier transform of the Bloch states:
\begin{align}
\ket{w_{n{\bf R}}}&=\frac{1}{\sqrt{N}}\sum_{\bf k}e^{-i{\bf k}\cdot {\bf R}}\ket{\psi_{n{\bf k}}}, \\
&=\frac{1}{\sqrt{N}}\sum_{\bf k}e^{i{\bf k}\cdot (\hat{\bf r}-{\bf R})}\ket{u_{n{\bf k}}}.
\end{align}
The Bloch states $\ket{\psi_{n{\bf k}}}$ are given by the inverse Fourier transform, given the Wannier states.
Because we assume that the Wannier functions are exponentially localized, its Bloch state is smooth over the whole Brillouin zone~\cite{brouder2007exponential}.

We first relate the representation of $I_{ST}$ symmetry in the Wannier basis and that in the Bloch basis. 
Since we are dealing with the case $(I_{ST})^2=+1$, we may take 
\begin{align}
\braket{ w_{\alpha,i, -{\bf R}+\Delta_{\alpha\beta}} |I_{ST} | w_{\beta,j ,{\bf R}}}=\delta_{ij}\delta_{\alpha,I_{ST}\beta},
\end{align}
with suitable unit cell translation $\Delta_{\alpha\beta}$. Here, $\alpha,\beta$ are Wyckoff position index and $i,j$ are orbital index (which, in fact, we do not really need for our purpose because when $I_{ST}$ symmetry is a site symmetry group element, its representation can be diagonalized in a spinless system). Also, $\Delta_{\alpha \beta}=-{\bf t}_\beta-{\bf t}_\alpha$ can easily be found from the action of $I_{ST}$ symmetry in real space, $-({\bf R}+{\bf t}_\beta)=-{\bf R}+\Delta_{\alpha \beta}+{\bf t}_\alpha$.

In the Bloch basis, 
\begin{align}
I_{ST} \ket{\psi_{\beta,j, {\bf k}}}=\frac{1}{\sqrt{N}}\sum_{\bf R} e^{-i\bf k\cdot {\bf R}}\ket{w_{\alpha,j,-{\bf R}+\Delta_{\alpha \beta}}}.
\end{align}
Thus, the sewing matrix is
\begin{equation}
\braket{ \psi_{\alpha,i, {\bf k}}|I_{ST}| \psi_{\beta,j, {\bf k}}}=e^{-i{\bf k}\cdot \Delta_{\alpha \beta}}\delta_{ij} \delta_{\alpha,I_{ST}\beta},
\end{equation}
where the constraint $-({\bf R}+{\bf t}_\beta)=-{\bf R}+\Delta_{\alpha \beta}+{\bf t}_\alpha$ or $\Delta_{\alpha \beta}=-{\bf t}_\alpha-{\bf t}_\beta$ is represented by $\delta_{\alpha,I_{ST} \beta}$. 

If we choose a real basis $\ket{\tilde{\psi}_{\alpha i{\bf k}}}=U_{\alpha i, \beta j }^\dagger ({\bf k})\ket{\psi_{\beta j {\bf k}}}$, the sewing matrix must be equal to $\delta_{\alpha \beta}$. 
The transformation above may not be continuous and periodic in general.
Our task now is to divide the Brillouin zone into two patches and determine whether the transition matrix for two bands resulting from two Wannier orbitals can have nontrivial winding.

The easiest case to consider is when the unit cell is left invariant under space-time inversion symmetry. This requires the space-time inversion symmetry to swap the position of the two Wannier centers. Then, $I_{ST}=\sigma_xK$ in the complex basis. Under the constant unitary transformation
\begin{align}
U=\frac{1}{\sqrt{2}}\begin{pmatrix}
e^{-3\pi i/4} & e^{3\pi i/4} \\
e^{3\pi i/4} & e^{-3\pi i/4}
\end{pmatrix},
\end{align}
$I_{ST}=K$ as required. Note that there is no need to introduce patches to define a real basis.

We next consider the case when the unit cell is not left invariant under the $I_{ST}$ symmetry, meaning that the Wannier centers occupy two different points invariant under inversion symmetry up to lattice translations. If we place the inversion center at one of the Wannier centers, the sewing matrix in the complex basis is 
\begin{align}
\begin{pmatrix}
1 & 0 \\
0 & e^{-i{\bf k}\cdot\Delta_{\alpha \beta}}
\end{pmatrix}.
\end{align}
We can obtain the real basis by taking 
\begin{align}
U=\begin{pmatrix}
1 & 0 \\
0 & e^{i{\bf k}\cdot\Delta_{\alpha \beta}/2}.
\end{pmatrix}
\end{align}
Without loss of generality, we can assume that $\Delta_{\alpha \beta}={\bf a}_1$, where $a_{1,2}$ are the two unit lattice vectors. 
Let us parametrize the Brillouin zone by $0\le k_1,k_2< 1$, where ${\bf k}=k_1{\bf G}_1+k_2{\bf G}_2$ and ${\bf G}_{1,2}$ are the reciprocal lattice vectors.
Then we introduce two patches, $N$ and $S$, covering $0\le k_1\le 1/2$ and $1/2 \le k_1\le 1\simeq 0$ respectively, and define $\ket{\tilde{\psi}^{N/S}_{\bf k}}=\ket{\tilde{\psi}_{\bf k}}$ for ${\bf k}\in N/S$. 
In the case of interest, 
the transition function is nontrivial only at $k_1=0$: $t^{NS}(0,k_y)=\braket{\tilde{\psi}^N(0,k_2)|\tilde{\psi}^S(0,k_2)}=\braket{\tilde{\psi}(0,k_2)|\tilde{\psi}(1,k_2)}=U^{-1}(0,k_2)U(1,k_2)=\sigma_z$. Since $\det \sigma_z=-1$, the real states are not orientable on the Brillouin zone, and the Euler class cannot not defined.
In conclusion, it is not possible to realize a nontrivial Euler class in the Brillouin zone with two exponentially localized $I_{ST}$-symmetric Wannier functions.

\section{Failure of Nielsen-Ninomiya Theorem due to the Euler class} 
\label{sec.Nielsen}

In the previous section, we have shown that the Euler class is a topological invariant characterizing the Wannier obstruction for two real bands.
Here we show that the Euler class is the topological invariant that explains the Wannier obstruction for nearly flat bands in TBG, which was attributed to the non-zero total winding number in the Brillouin zone. More explicitly, we show that the Euler class is equivalent to half the total winding number.
To introduce some notations and set the stage for the discussion that follows, we first give a short proof of the 2D Nielsen-Ninomiya theorem, in analogy to the three-dimensional (3D) case~\cite{witten2015three}. 
Our main result will follow by carefully investigating the failure of the 2D Nielsen-Ninomiya theorem.

\subsection{Two-dimensional Nielsen-Ninomiya theorem} \label{Nielsen-proof}

In this section, we give a short proof of the 2D Nielsen-Ninomiya theorem that the total winding number is zero in 2D periodic systems and point out what the assumptions are. Note that we have stated this theorem by using the winding number instead of Berry phase because Berry phase is defined only modulo $2\pi$.

Let us take two real basis states $\ket{\tilde{u}_{1\bf k}}$ and $\ket{\tilde{u}_{2\bf k}}$ such that $I_{ST}$ is represented by the complex conjugation $K$ (i.e., $I_{ST}=K$), so the $I_{ST}$ symmetry condition $I_{ST}H({\bf k})(I_{ST})^{-1}=H({\bf k})$ requires that the matrix elements of the Hamiltonian $H_{mn}({\bf k})=\braket{\tilde{u}_{m\bf k}|H({\bf k})|\tilde{u}_{n\bf k}}$ to be real, that is, $H_{mn}({\bf k})=H_{mn}^*({\bf k})$.
Therefore,
\begin{equation}
\label{eq.2-bandH}
H({\bf k})=r({\bf k})\cos \theta({\bf k})\sigma_1+r({\bf k})\sin \theta({\bf k})\sigma_3
\end{equation}
where $r({\bf k})\ge 0$, $\sigma_1$ and $\sigma_3$ are Pauli matrices defined in the basis $\{\ket{\tilde{u}_{1\bf k}},\ket{\tilde{u}_{2\bf k}}\}$, and a term proportional to $\sigma_0$ is ignored. Let us define a unit vector ${\bf n}({\bf k})=(\cos \theta({\bf k}), \sin \theta({\bf k}))$ away from points at which $r(\bf k)=0$. The winding number of the Hamiltonian along a loop $C$ is defined to be the winding number of ${\bf n}({\bf k})$~\cite{chiu2016classification}:$N_C=\frac{1}{2\pi}\oint_C d{\bf k}\cdot \nabla_{\bf k}\theta({\bf k})$.
Let $D_i$ be a disk enclosing an $i$th vortex, so that the total winding number is given by
\begin{equation}
N_t=\frac{1}{2\pi} \oint_{\cup_i \partial D_i } d{\bf k}\cdot \nabla_{\bf k}\theta({\bf k}),
\end{equation}
where $\partial D_i$ is the boundary of $D_{i}$.
Using the Stokes' theorem, we have
\begin{equation}
N_t=-\int_{BZ-\cup_i D_i}d{\bf S}\cdot \nabla_{\bf k}\times \nabla_{\bf k}\theta({\bf k})=0
\end{equation}

Here, we have made an obvious assumption that the matrix elements of the two-band Hamiltonian are continuously defined throughout the Brillouin zone. This has two important implications. The first one is that when the matrix elements of the two-band Hamiltonian cannot be defined continuously in the presence of $I_{ST}$ symmetry, a non-vanishing total winding number is allowed. We will discuss this in the following subsection. The second implication is that when the two bands are no longer isolated from the other bands, the winding number may lose its meaning. This will be discussed further in section~\ref{subsec.vortex_annihilation}.

\subsection{Winding number and the Euler class} 
\label{subsec.winding=Euler}

\begin{figure}[t!]
\includegraphics[width=7cm]{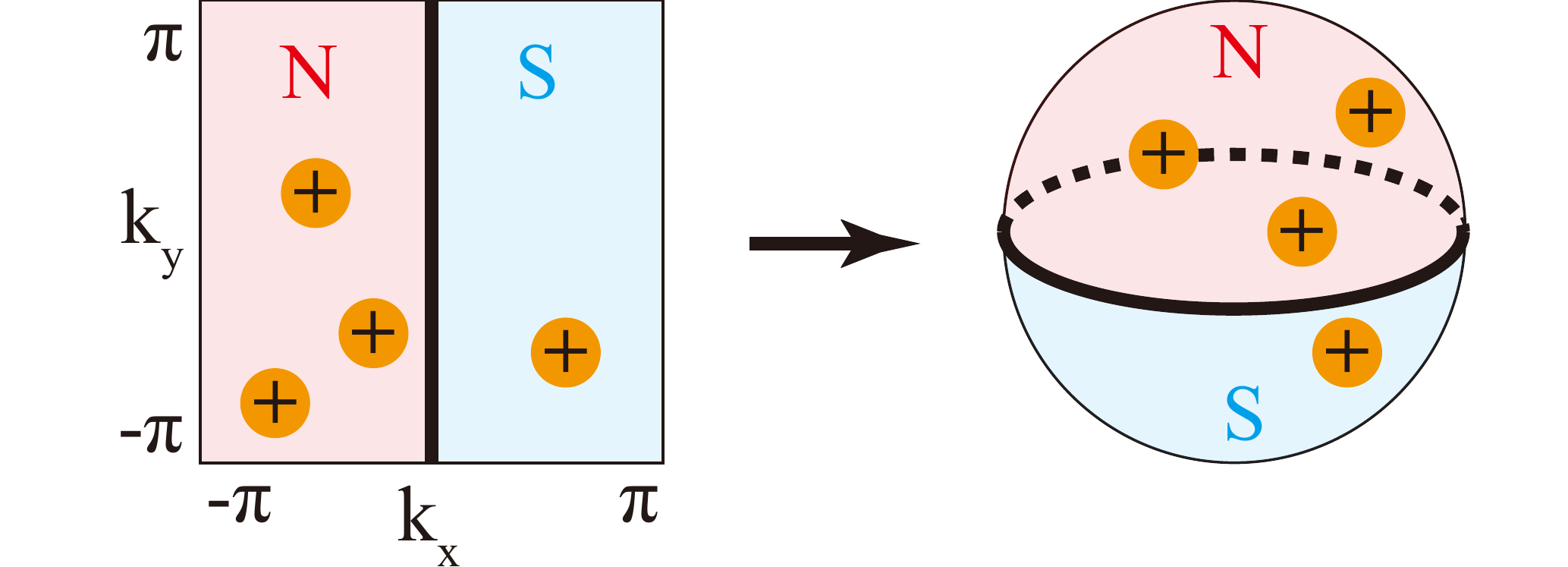}
\caption{Deformation of the Brillouin zone to a sphere.
When the total winding number is non-zero in the Brillouin zone, Hamiltonian matrix elements are smooth only over local patches $N$ and $S$, respectively.
When the nontrivial transitions are restricted to $k_x=0$ (the black bold line), the boundary of the Brillouin zone can be contracted to a point so that the Brillouin zone becomes a sphere on the right.}
\label{fig.patch}
\end{figure}

Let us now prove that $e_2$ is equal to half the total winding number of a two band Hamiltonian. 
We again consider the Hamiltonian in Eq.~\eqref{eq.2-bandH}. 

In the case when the total Berry phase, i.e. the sum of the Berry phases of the two bands, along any non-contractible 1D cycle in the Brillouin zone is trivial, we can take a spherical gauge in which we neglect the non-contractible 1D cycles and instead view the Brillouin zone as a sphere [Fig.~\ref{fig.patch}].
We will discuss the case in which the Berry phase is nontrivial in Sec.~\ref{subsec.non-orientable}

One immediate consequence of the non-vanishing total winding number is that it is impossible to define a continuous Hamiltonian matrix elements throughout the sphere. 
Thus, let us divide the sphere into $N$ and $S$ hemispheres such that each vortex is located in the interior of either the $N$ or $S$ hemisphere [Fig.~\ref{fig.patch}]. On the equator, we need a transition function, $O_{NS}(\phi)\in SO(2)$, where $\phi$  is the azimuthal angle parametrizing the equator. The two Hamiltonians on the $N$ and $S$ hemispheres are connected along the equator as 
\begin{equation}
\label{eq.transitionH}
(H_N)_{mn}=(O_{NS})_{mp} (H_S)_{pq} (O_{NS}^\dagger)_{qn}
\end{equation}
Thus, we must have $O_{NS}=\exp(-i\sigma_y(\theta_{S}-\theta_{N})/2)$.
Before moving on, note that we may assume that the two bands of our interest arise as sub-bands of a lattice Hamiltonian. Then, this transition matrix is the transition function between the two sub-bands of interest. Because the full lattice Hamiltonian is continuous, any discontinuity of the projected $2\times 2$ Hamiltonian must originate from that of the basis states of the two subbands.  
Accordingly, the Euler class, which is given by the winding number of the transition function, is equal to 
\begin{align}
\frac{1}{4\pi}\oint_{\rm equator}d{\bf k}\cdot (\nabla_{\bf k}\theta^N-\nabla_{\bf k}\theta^S)=(N_N+N_S)/2, 
\end{align}
where $N_{N/S}$ are the sum of the winding number within $N/S$ patch. 
The negative sign in the definition of $N_S$ is there because the winding number is defined by the counterclockwise line integral with respect to the normal direction of the sphere.
In conclusion, we have proved that
\begin{align}
\label{eq.Euler=winding}
e_2=-\frac{1}{2}N_t.
\end{align}
Let us note that this is a generalization of the Poincar\'e-Hopf theorem~\cite{mathai2017differential,bott2013differential,choquet1982analysis}, which relates zeros of a tangent vector field to the Euler characteristic of the manifold, to rank two real Bloch bundles (i.e., two real Bloch states).

\section{Off-diagonal Berry phase}
\label{sec.off-diagonal}
The relation in Eq.~\eqref{eq.Euler=winding} allows us to study the Euler class by investigating band degeneracies which carry nontrivial winding numbers.
However, it is not easy to treat the winding number with its conventional definition, because it requires nontrivial transition functions between local patches when the total winding number in the Brillouin zone is non-zero.
Instead of using the matrix element of the Hamiltonian, here we develop a new method for calculating the winding number of vortices by using energy eigenstates.
We will show that the winding number of a vortex can be calculated by using an off-diagonal component of the Berry connection.
Although we focus on $I_{ST}$-symmetric two-band systems here, the same method can be applied to any chiral symmetric system [See Appdendix~\ref{sec.chiral} for details].
Moreover, since the energy eigenstates can be taken to be smooth everywhere on the Brillouin zone except at the points of degeneracy under a smooth complex gauge, the off-diagonal Berry connection can also be smoothly defined on the punctured Brillouin zone without the need of introducing patches.
Because of this reason, in this section, we relax the reality condition $I_{ST}\ket{\tilde{u}_{n\bf k}}=\ket{\tilde{u}_{n\bf k}}$, and instead use a smooth complex gauge to define the off-diagonal Berry phase.
This method will be particularly useful when we study pair annihilations of vortices in Sec.~\ref{sec.nonorientable}.

\subsection{Sewing matrix and the Berry connection}

Let $\{\ket{u_{n\bf k}}\}$ be energy eigenstates with energy $E_{n\bf k}$.
In this basis, the sewing matrix $G$ of the $I_{ST}$ operator is defined by
\begin{align}
\label{eq.PT-sewing}
G_{mn}({\bf k})=\braket{u_{m\bf k}|I_{ST}|u_{n\bf k}}.
\end{align}
This sewing matrix is diagonal when the energy eigenstates are non-degenerate, because $I_{ST}$ operator does not change the energy of the state when it is a symmetry operator.
Then
\begin{align}
G({\bf k})
&=
\begin{pmatrix}
e^{i\theta_1({\bf k})}&0\\
0&e^{i\theta_2({\bf k})}
\end{pmatrix}.
\end{align}

Equation~\eqref{eq.PT-sewing} can be used to show that the Berry connection 
\begin{align}
A_{mn}({\bf k})=\braket{u_{m\bf k}|\nabla_{\bf k}|u_{n\bf k}},
\end{align}
in $I_{ST}$-symmetric systems satisfies
\begin{align}
{\bf A}({\bf k})=G({\bf k}){\bf A}^*({\bf k})G^{-1}({\bf k})+G({\bf k})\nabla_{\bf k}G^{-1}({\bf k}).
\end{align}
In a two-band system, or more generally for two subbands of a larger system, the constraint equation can be exactly solved on the non-degenerate region.
We have
\begin{align}
{\bf A}({\bf k})
&=
\begin{pmatrix}
-\frac{i}{2}\nabla_{\bf k}\theta_1({\bf k})&{\bf a}({\bf k})e^{i\chi({\bf k})}\\
-{\bf a}({\bf k})e^{-i\chi({\bf k})}&-\frac{i}{2}\nabla_{\bf k}\theta_2({\bf k})
\end{pmatrix},
\end{align}
where $\chi({\bf k})=(\theta_1({\bf k})-\theta_2({\bf k}))/2$, and we defined ${\bf a}({\bf k})= e^{-i\chi({\bf k})}{\bf A}_{12}({\bf k})$, which is the only real parameter undetermined by the sewing matrix.
Here $\chi({\bf k})$ is defined modulo $\pi$ because $\theta_1({\bf k})$ and $\theta_2({\bf k})$ are defined modulo $2\pi$.
Correspondingly, a definite global sign of ${\bf a}({\bf k})$ is fixed after choosing the global phase of $e^{i\chi({\bf k})}$.

Let us emphasize that ${\bf a}({\bf k})$ is the gauge-invariant part of the off-diagonal Berry connection ${\bf A}_{12}({\bf k})$: it is invariant under diagonal gauge transformations, which do not mix different energy eigenstates.
Under a gauge transformation 
\begin{align}
\ket{u_{n\bf k}}\rightarrow \ket{u'_{n\bf k}}=e^{i\zeta_{n}({\bf k})}\ket{u_{n\bf k}},
\end{align}
where $n=1,2$, we have
$
\theta_{n}'({\bf k})
=\theta_{n}({\bf k})-2\zeta_{n}({\bf k})$, and
$
{\bf A}'_{12}({\bf k})
=e^{-i(\zeta_1({\bf k})-\zeta_2({\bf k}))}{\bf A}_{12}({\bf k}).
$
Then
\begin{align}
{\bf a}'({\bf k})
={\bf A}'_{12}({\bf k})e^{-i\chi'({\bf k})}
={\bf A}_{12}({\bf k})e^{-i\chi({\bf k})}
={\bf a}({\bf k}).
\end{align}

\subsection{Winding number from the off-diagonal Berry connection}

Now we show that ${\bf a}({\bf k})$ contains the full information on the winding number.
Let us consider the following eigenstates of the two-band Hamiltonian in Eq.~\eqref{eq.2-bandH}.
\begin{align}
\ket{u_{1\bf k}}
=
\begin{pmatrix}
\sin \phi({\bf k})\\
\cos \phi({\bf k})
\end{pmatrix},\quad
\ket{u_{2\bf k}}
=
\begin{pmatrix}
-\cos \phi({\bf k})\\
\sin\phi({\bf k})
\end{pmatrix},
\end{align}
where $\phi({\bf k})=\theta({\bf k})/2-\pi/4$.
In this choice of gauge, $G({\bf k})=1$, and the Berry connection is given by
\begin{align}
{\bf A}({\bf k})
&=
\begin{pmatrix}
0&\frac{1}{2}\nabla_{\bf k}\theta({\bf k})\\
-\frac{1}{2}\nabla_{\bf k}\theta({\bf k})&0
\end{pmatrix}.
\end{align}
From this expression, we get
$
{\bf a}({\bf k})=\frac{1}{2}\nabla_{\bf k}\theta({\bf k}),
$
such that
\begin{align}
\label{eq.ODBP=winding}
\oint_{S^1} d{\bf k}\cdot {\bf a}({\bf k})=\frac{1}{2}\oint_{S^1} d{\bf k}\cdot \nabla_{\bf k}\theta({\bf k})=N_{S^1}\pi.
\end{align}
Since ${\bf a}({\bf k})$ is invariant under any diagonal gauge transformations, the {\it off-diagonal Berry phase} defined by $\oint_{S^1} d{\bf k}\cdot {\bf a}({\bf k})$ in any smooth energy eigenstate basis gives the desired winding number $N_{S^1}$.

When we consider two subbands of a larger system, the off-diagonal Berry phase can still capture the winding number of vortices although it is not quantized in general.
As one can see from $\nabla_{\bf k}\times {\bf a}({\bf k})=\tilde{F}_{12}\ne 0$ in a real eigenstate basis, $\oint_{S^1} d{\bf k}\cdot {\bf a}({\bf k})$ is not quantized.
However, the above relation between the off-diagonal Berry phase and the winding number in Eq.~\eqref{eq.ODBP=winding} is still valid in the vicinity of a vortex, where the other bands except the two bands of our interest contribute to the off-diagonal Berry phase negligibly.
In other words, as a disk $D$ containing a vortex $v$ shrinks to the vortex site, we have
\begin{align}
\oint_{\partial D\rightarrow v} d{\bf k}\cdot {\bf a}({\bf k})=N(v)\pi,
\end{align}
where $N(v)$ is the winding number of a vortex $v$.
This is consistent with the correspondence between the Euler class and the winding number we derived above.
Consider a punctured sphere $S^2_{\text{p}}\equiv S^2-\sum_iD_i$, where $D_i$ is an infinitesimal disk on the sphere containing a vortex $v_i$. Then, in the limit of vanishing $D_i$, we find
\begin{align}
e_2
&=\frac{1}{2\pi}\oint_{S^2_{\text{p}}}d{\bf S}\cdot \nabla_{\bf k}\times {\bf a}({\bf k})
\nonumber\\
&=                                                                                                                                                                                                                                                                                                                                                                                                                                                                                                                         - \frac{1}{2\pi}\sum_i\oint_{\d D_i}d{\bf k}\cdot {\bf a}({\bf k})
\nonumber\\
&=-\frac{1}{2}N_t.
\end{align}

\section{Pair annihilation of vortices}
\label{sec.nonorientable}

\begin{figure}[t!]
\includegraphics[width=8.5cm]{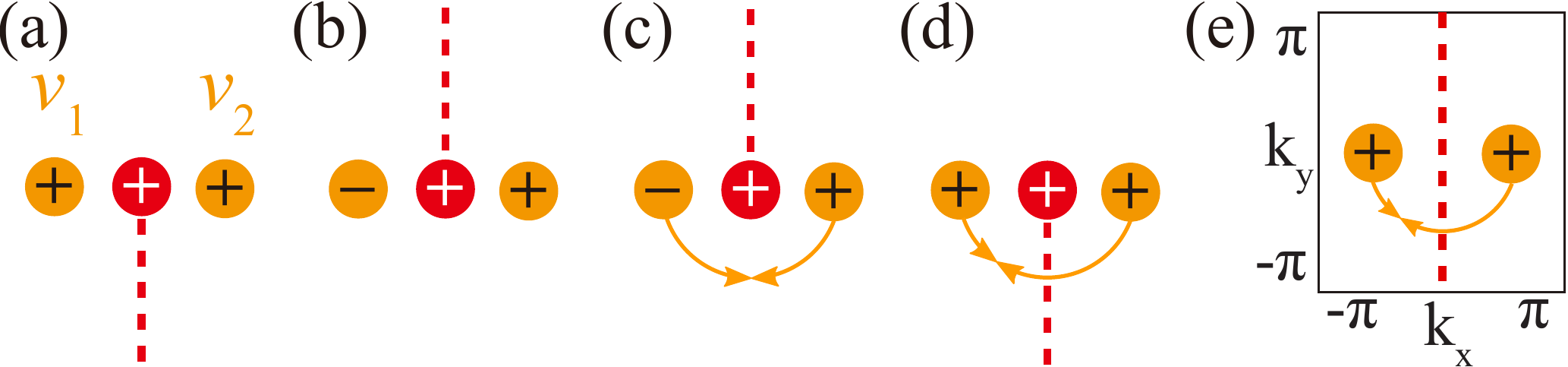}
\caption{Pair annihilation process in a 2D Brillouin zone. (a) $v_1$ and $v_2$ are the vortices between bands 1 and 2. They are indicated by orange dots with the $+$ or $-$ representing their winding number. The red dot indicates a $\pi$ Berry flux generator, which can be thought of as a vortex between bands 2 and 3. The Dirac string is represented by the red dashed line. (b) An alternative choice of the Dirac string. (c) The vortices can only be annihilated by moving downwards. (d) A vortex reverses its winding number when it crosses a Dirac string. Thus, the vortices can be annihilated as shown.
(e) Pair annihilation of two vortices with the same winding number in a Brillouin zone. A Dirac string along the $k_y$ direction indicates the non-trivial Berry phase along the $k_x$ direction, which allows the pair annihilation of the two vortices.
}
\label{fig.annihilation}
\end{figure}

In this section, we discuss how a pair annihilation of vortices can occur. In the previous sections, we have described how a non-zero Euler class gives a non-zero total winding number, and how the winding number can be defined in terms of the off-diagonal Berry connection. A crucial assumption for achieving a non-zero total winding number was that the total Berry phase along any non-contractible cycle must be zero.

To study the effect of non-zero Berry phase, let us notice that it is impossible to consistently choose a definite global sign of ${\bf a}({\bf k})$ when the total Berry phase is nontrivial along a loop.
Suppose we take a smooth and periodic gauge around a loop $C$ parametrized by $0\le k<2\pi$.
Then, the sewing matrix and the Berry connection are also smooth and periodic along the cycle. 
The periodic condition $G(2\pi)=G(0)$ gives
\begin{align}
e^{i\chi(2\pi)}
&=e^{i(\theta_1(2\pi)-\theta_1(0)-[\theta_2(2\pi)-\theta_2(0)])/2}e^{i\chi(0)}
\nonumber\\
&=e^{i(\theta_1(2\pi)-\theta_1(0)+\theta_2(2\pi)-\theta_2(0))/2}e^{i\chi(0)}
\nonumber\\
&=e^{-\oint_Cd{\bf k}\cdot {\rm Tr}{\bf A}}e^{i\chi(0)}.
\end{align}
Since ${\bf A}(2\pi)={\bf A}(0)$, we find that
\begin{align}
\label{eq.periodicity}
{\bf a}(2\pi)
=e^{\oint_Cd{\bf k}\cdot {\rm Tr}{\bf A}}{\bf a}(0).
\end{align}
Thus, we cannot assign the global sign of ${\bf a}({\bf k})$ unambiguously when the total Berry phase is nontrivial.
This implies that a pair creation of two vortices with the same winding number can occur when the band gap closes and the nontrivial Berry phase is generated by gap-closing points.
In this section, we describe this mechanism of the pair creation and annihilation of vortices.
We also comment on the case with nontrivial Berry phase along the non-contractible 1D cycles in the Brillouin zone.

\subsection{Pair annihilation process}
\label{subsec.vortex_annihilation}

In Fig.~\ref{fig.annihilation} (a), we show a schematic picture of a part of the 2D Brillouin zone.
The orange dots labeled by $v_1$ and $v_2$ represent two vortices between energy bands 1 and 2, which are the bands we are interested in. Let us assume that $v_1$ and $v_2$ have the same winding number. We will describe how $v_1$ and $v_2$ can be pair-annihilated when the band gap between these two bands and another band (band 3) closes to form additional gap closing points (Dirac points). Notice that in the viewpoint of bands 1 and 2, such an additional gap closing point acts as a $\pi$ Berry phase generator in the sense that the sum of the Berry phases for bands 1 and 2 calculated around a loop enclosing the additional gap closing point formed by bands 1 or 2 and the band 3 is $\pi$. Such a $\pi$ Berry phase generator is shown as a red dot in Fig.~\ref{fig.annihilation}.

According to Eq.~\eqref{eq.periodicity}, ${\bf a}({\bf k})$ changes sign when it circles around the red dot once, because of the $\pi$ Berry phase.
For the purpose of discussing the winding number of vortices, we must therefore introduce a branch cut, shown as a dashed line in Fig.~\ref{fig.annihilation} (a). 
Across this branch cut, the sign of both ${\bf a}({\bf k})$ and $e^{i \chi({\bf k})}$ changes, so that ${\bf A}_{12}$ is well defined.
We will refer to this branch cut as a {\it a Dirac string}, in analogy to the Dirac string that arises from three-dimensional magnetic monopoles~\cite{dirac1931quantised}.
As in the three-dimensional case, this Dirac string also ends when it reaches another $\pi$ Berry flux generator because the total Berry phase surrounding the two $\pi$ Berry flux generator is $2\pi$ so that the factor $e^{i\chi({\bf k})}$ is well defined around any curve surrounding them.  
 
To illustrate the most important property of the Dirac string, suppose that $v_1$ and $v_2$ have the same winding number with the choice of the Dirac string in Fig.~\ref{fig.annihilation}(a).
Then, consider a process in which the Dirac string rotates clockwise to the configuration shown in Fig.~\ref{fig.annihilation} (b).
This is equivalent to changing the sign of ${\bf a}({\bf k})$ at the points where the Dirac string sweeps, so that the winding number of $v_1$ also changes.
One implication of this result is that $v_1$ and $v_2$ can be annihilated only by circling around the red dot downwards, as shown in Fig.~\ref{fig.annihilation} (c).
Also, by considering the reverse process in which the Dirac string is fixed and the vortices move, one sees that \textit{whenever a vortex crosses a Dirac string, its winding number changes its sign}.
Thus, if we consider the annihilation process shown in Fig.~\ref{fig.annihilation} (d), the winding number of the vortex $v_2$ changes the sign upon crossing the Dirac string, before $v_1$ and $v_2$ are pair annihilated.

\subsection{Instability of vortices in non-orientable cases} 
\label{subsec.non-orientable}

Up to now, we have dealt with the case when the Euler class is well defined by assuming that the total Berry phase along any non-contractible cycle in the Brillouin zone is trivial.
However, when there is a nontrivial Berry phase along any non-contractible 1D cycle on the Brillouin zone torus, and thus the Euler class is ill-defined, two vortices with the same winding number can be pair annihilated even when band 1 and 2 are well separated from other bands.
The reason why two vortices can be pair-annihilated is basically the same as the previous case discussed in section \ref{subsec.vortex_annihilation}.
Namely, the nontrivial Berry phase along a nontrivial cycle implies that there must be a closed Dirac string along the other non-contractible 1D cycle of the Brillouin zone torus.
Because the winding number of a vortex changes whenever it crosses a Dirac string, even if two vortices have the same winding number at the beginning, after transporting one of the vortices across the Dirac string, two vortices can be pair-annihilated as shown in Fig.~\ref{fig.annihilation} (e).

\begin{figure}[t!]
\includegraphics[width=8.5cm]{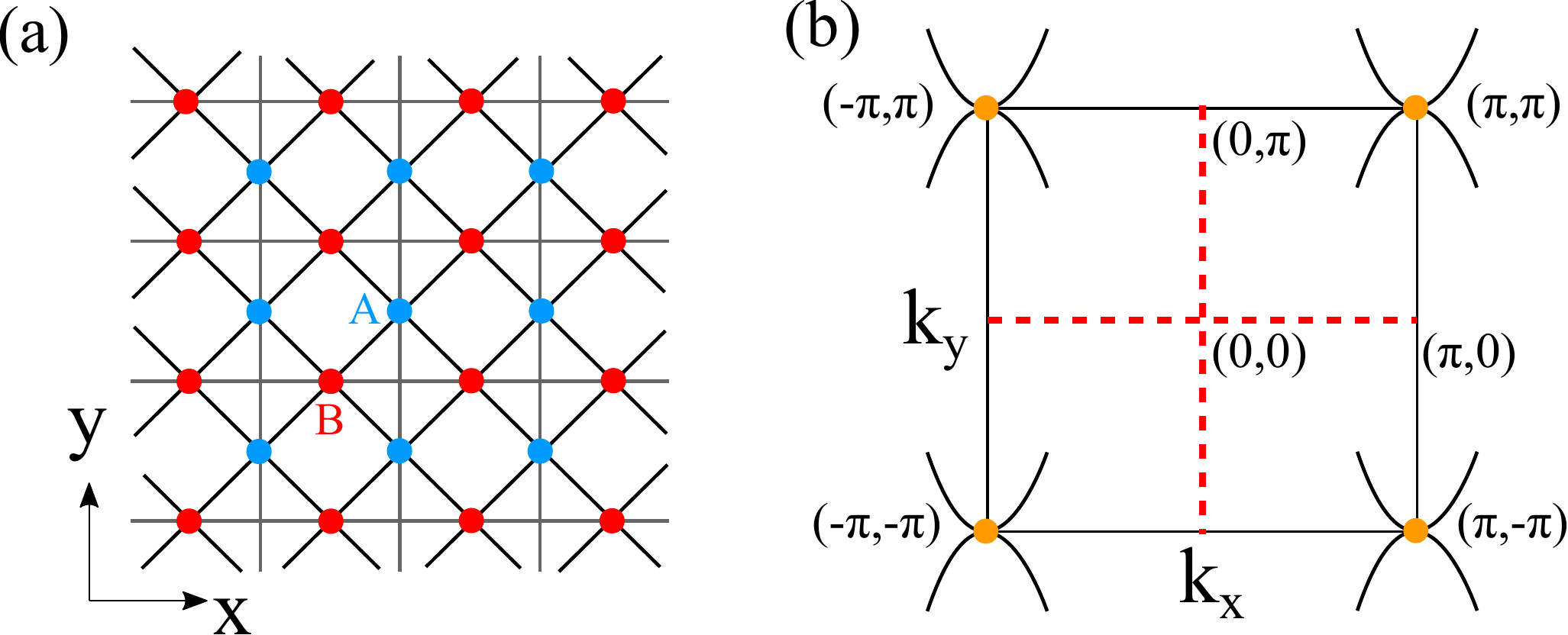}
\caption{Quadratic band crossing model introduced by Sun et. al.,\cite{sun2009topological}.
(a) The structure of the checkerboard lattice with two sites in a unit cell where the blue (red) dot represents A (B) sublattice site. (b) Low energy band structure in the Brillouin zone. 
The orange dot at $M=(\pi,\pi)$ represents a quadratic band crossing with the winding number $\pm2$. A red dotted line indicates a Dirac string across which the winding number of a band crossing point changes the sign.  
}
\label{fig.checkerboard}
\end{figure}

For instance, let us consider the two-band lattice model on checkerboard lattice proposed in~\cite{sun2009topological}, which falls exactly into this category.
As explained below, this model contains a single quadratic band crossing point (QBCP) with the winding number $\pm 2$ at $M=(\pi,\pi)$ in the BZ. 
The presence of a well-defined tight-binding Hamiltonian indicates that there is no Wannier obstruction for the two bands, and thus the Euler class of this model should be zero. Naively, the presence of the band crossing point with the winding number two seems to be incompatible with the fact that the Euler class of the system is trivial. One way to reconcile this contradiction is to consider nontrivial Berry phase along non-contractible cycles of the BZ. Below we show that this is indeed the case, that is, the total winding number is ill-defined due to the $\pi$ Berry phase along non-contractible cycles of the BZ.

The checkerboard lattice is shown in Fig.~\ref{fig.checkerboard}(a). The relevant tight binding Hamiltonian with one orbital per lattice site is
\begin{equation}
H=-\sum_{ij} t_{ij} c^\dagger_i c_j,
\end{equation}
where $t_{ij}=t$ for nearest neighbor sites, $t_{ij}=t' (t'')$ for next nearest neighbor sites connected (not connected) by vertical or horizontal bonds. This spinless model has time reversal $T$ symmetry and four-fold rotation $C_{4z}$ symmetry about the center of the smallest square formed by $A$ (blue) and $B$ (red) sites. Since the system has $C_{2z}T=(C_{4z})^{2}T$ symmetry, the theoretical idea developed in the preceding sections can be directly applied. 

After we Fourier transform the Hamiltonian by taking into account the atomic positions within the unit cell, we obtain
\begin{equation}
H({\bf k})=d_0({\bf k})\sigma_0+d_x({\bf k}) \sigma_x+d_z({\bf k}) \sigma_z,
\end{equation}
where $d_0({\bf k})=-(t'+t'') (\cos k_x+\cos k_y)$, $d_x({\bf k})=-4 t \cos \frac{k_x}{2} \cos \frac{k_y}{2}$, and $d_z({\bf k})=-(t'-t'')(\cos k_y-\cos k_x)$. 
It is important to note that this Hamiltonian is real but not periodic. 
In contrast, if we take the Fourier transformation with respect to the position of the unit cell neglecting the atomic positions in the unit cell, $d_0({\bf k})$ and $d_z({\bf k})$ remain the same, but we now have $d_x({\bf k})-i d_y({\bf k})=-(1+e^{-ik_x}+e^{-ik_y}+e^{-i(k_x+k_y)})$. 
Thus, the Hamiltonian is complex and periodic in this case. 

If we choose the real basis in which the winding number is well-defined and expand the Hamiltonian near the $M$ point, we obtain $d_x=-tk_x k_y$ and $d_z=\frac{(t'-t'')}{2}(k_x^2-k_y^2)$ so that the winding number is $\pm 2$ where the sign depends on the choice of the parameters.
We find $H({\bf k}+{\bf G}_i)=\sigma_z H({\bf k}) \sigma_z$, where ${\bf G}_i$ is the reciprocal lattice vector either along the $k_x$ or $k_y$ direction. This indicates that the Hamiltonian is discontinuous at the BZ boundary. Since $\det (\sigma_z)=-1$, an orientation reversing transformation is necessary to glue the Hamiltonian matrix elements at the BZ boundary.

This non-orientability indicates that the total Berry phase along the $k_x$ and $k_y$ directions should be $\pi$, which can be explicitly checked by computing the Berry phase using a complex smooth basis.

As shown in Sec.~\ref{subsec.vortex_annihilation}, the $\pi$-Berry phase along both the $k_x$ and $k_y$ directions indicates the presence of Dirac strings along the two non-contractible cycles of the Brillouin zone. [See Fig.~\ref{fig.checkerboard}(b).] If the $C_{4z}$ symmetry is broken while $C_{2z}$ is preserved, the QBCP can be split into two Dirac points and be annihilated when they merge at $X=(\pi,0)$ or $Y=(0,\pi)$ after crossing a Dirac string. This phenomenon is indeed observed in a related tight binding model on the checkerboard lattice in \cite{montambaux2018winding}.

Let us note that the appearance of the Dirac string is related to the absence of a $C_{2z}$-invariant unit cell. 
If we Fourier transform a tight-binding Hamiltonian, we have $H({\bf k}+{\bf G})=V^{-1}({\bf G})H({\bf k})V({\bf G})$ in general, where $V_{\alpha\beta}({\bf k})=\exp(i{\bf k}\cdot {\bf r}_\alpha)\delta_{\alpha\beta}$, and $\alpha,\beta$ are indices labelling the atomic sites.
When $\det V({\bf G})=-1$, an odd number of atoms are displaced by a half lattice vector from the $C_{2z}$ center, so the corresponding unit cell is not $C_{2z}$-invariant.

\subsection{Topological phase transition} 
\label{subsec.tpt}

\begin{figure}[t!]
\includegraphics[width=8.5cm]{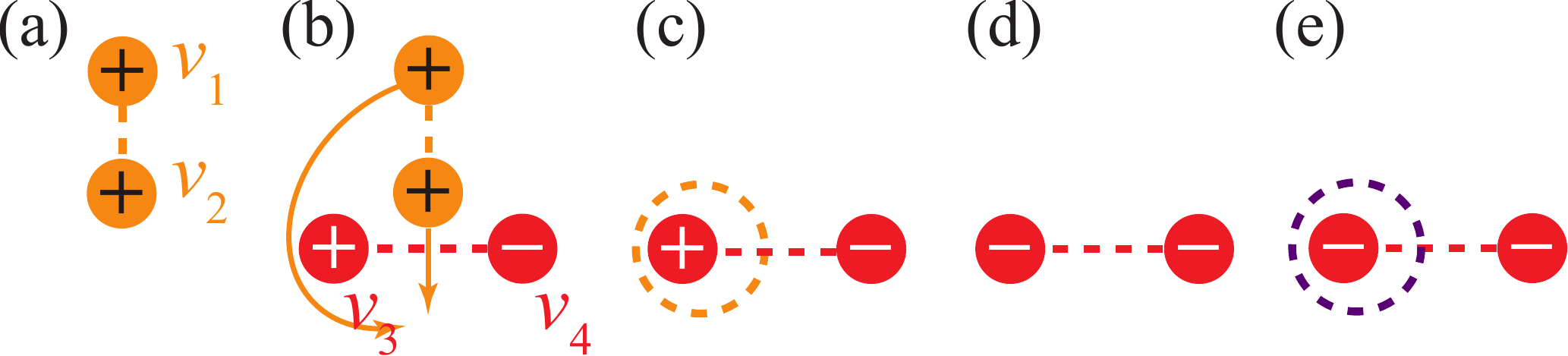}
\caption{Topological phase transition from $e_2=1$ to $e_2=0$ by a pair annihilation of vortices. 
(a) $v_1$ and $v_2$ are the vortices formed by two occupied bands (bands 1 and 2). The orange dashed line is the Dirac string of $v_1$ and $v_2$ in the viewpoint of bands 2 and 3 where band 3 is the lowest energy unoccupied band.
(b) $v_3$ and $v_4$ are vortices that form when the band gap between the occupied (bands 1, 2) and unoccupied band (band 3) closes. The red dashed line is the Dirac string between $v_3$ and $v_4$ as seen by bands 1 and 2. $v_1$ and $v_2$ can be annihilated as shown, because $v_2$ changes its winding number when it crosses the red dashed line.
(c)  Annihilation of $v_1$ and $v_2$ leaves behind a Dirac string as shown.
(d) Due to the Dirac string left behind in (c), the winding number of $v_3$ changes.
(e) It may seem that $v_3$ and $v_4$ cannot annihilate, but this is not the case because processes similar to those shown in (a)$\sim$(c) occur in the unoccupied bands to leave another loop of a Dirac string (purple), which also changes the winding number of $v_3$.
}
\label{fig.tpt1}
\end{figure}

Let us now explain how the vortex annihilation can be used to describe the topological phase transition from a $e_2=1$ phase to a $e_2=0$ phase.
For a minimal description, we consider a four-band system at half-filling, where the occupied bands (band 1,2) have $e_2=1$, and the unoccupied bands (band 3,4) have $e_2=-1$ as in the case shown in Fig.~\ref{fig.TB}. 
Recalling that an insulator with $|e_2|=1$ has a pair of vortices with the same winding number, we must either annihilate the two vortices or create another pair of vortices with the opposite winding number so that the total winding number of bands 1 and 2 becomes zero.
For simplicity, we discuss only the former case, shown in Fig.~\ref{fig.tpt1} (a)$\sim$(e). 

Following the convention in the previous section, the pair of vortices between bands 1 and 2 with the same winding numbers are labeled by $v_1$ and $v_2$, as shown in Fig.~\ref{fig.tpt1} (a). 
For the phase transition to occur, a pair of vortices with opposite winding number ( $v_3$ and $v_4$) must be formed between bands 2 and 3 via a band gap closing, as shown in Fig.~\ref{fig.tpt1}(b). 
Notice that we have drawn two Dirac strings for each of the pairs, because in the viewpoint of bands 1 and 2, $v_3$ and $v_4$ act as $\pi$ Berry flux generators, while in the viewpoint of bands 2 and 3, $v_1$ and $v_2$ act as $\pi$ Berry flux generators.
Thus, $v_1$ and $v_2$ can be annihilated by passing through the Dirac string, as shown in Fig.~\ref{fig.tpt1} (b) and (c). 
However, this leaves behind a ring of the Dirac string, as shown in Fig.~\ref{fig.tpt1} (c), so that $v_3$ and $v_4$ eventually have the same winding number as in Fig.~\ref{fig.tpt1} (d) after $v_3$ crosses the ring of the Dirac string which shrinks to a point and disappears in the end. However, we have only focused on bands 1, 2, and 3, but we must not forget about the vortices between bands 3 and 4. When the vortices in bands 3 and 4 go through a similar annihilation process, another Dirac string forming a ring will be left as in Fig.~\ref{fig.tpt1} (c). This will in turn change the winding number of $v_3$ or $v_4$ in Fig.~\ref{fig.tpt1} (e). Thus, $v_3$ and $v_4$ can be annihilated to open up the band gap, resulting in a trivial insulator.

\section{Fragile topology and higher-order topology}
\label{sec.fragile}

\begin{figure}[t!]
\includegraphics[width=8.5cm]{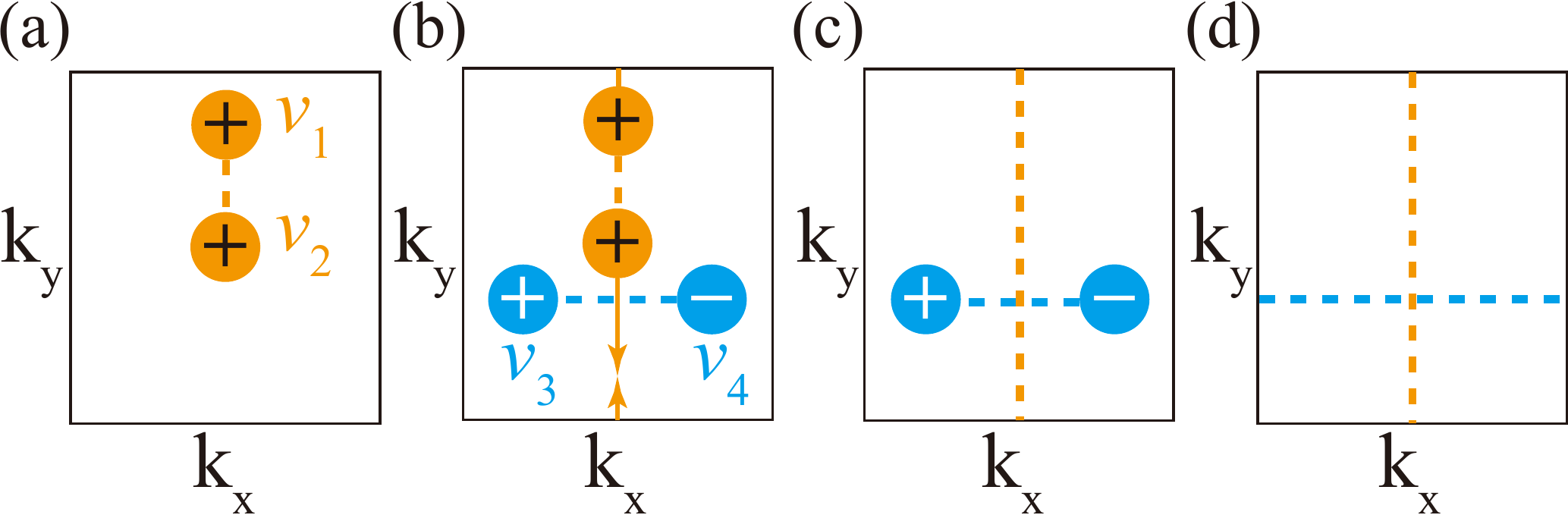}
\caption{
Fragility of vortices in a two-band system (bands 1, 2) with $e_2=1$ against adding one trivial band (band 0) below the Fermi energy.
The added trivial band is assumed to have the lowest energy level.
The box represents the 2D Brillouin zone.
(a) Vortices $v_1$ and $v_2$ with the same winding number formed between band 1 and band 2. The dashed orange line is the Dirac string for vortices which may be formed between band 0 and band 1.
(b) Pair annihilation of $v_1$ and $v_2$ after a band inversion between band 0 and band 1. Blue vortices $v_3$ and $v_4$ are pair-created after the band inversion between bands 0 and 1.
$v_1$ and $v_2$ can be pair-annihilated because the winding number of $v_2$ changes the sign after it crosses the blue Dirac string.
(c) The orange Dirac string extends along a non-contractible cycle after $v_1$ and $v_2$ are pair-annihilated.
(d) The blue Dirac string also winds a non-contractible 1D cycle after $v_3$ and $v_4$ are pair-annihilated.
}
\label{fig.fragile}
\end{figure}

In Ref.~\cite{po2018origin}, Po et al. have conjectured that the topological characteristic of two bands having two vortices with the identical winding number is fragile against adding topologically trivial bands~\cite{po2017fragile}, based on the observation that the integer winding number of the vortices is defined only for two bands.
Our theory is consistent with this conjecture in that the Euler class is also defined only for two bands.
However, there is a caveat.
Although the Euler class is defined only for two bands, its parity remains meaningful even when the number of bands becomes larger than two due to the additional trivial bands. In fact, the Euler class modulo two is identical to another $Z_{2}$ topological invariant, known as second Stiefel-Whitney class $w_2$, that is well-defined for any number of bands. Namely, if the Euler class of the two-band model is even (odd), $w_{2}$ of the system should remain zero (one) after the inclusion of additional trivial bands~\cite{ahn2018linking}.
Such a change of the topological indices from $Z$ to $Z_{2}$ can be observed from the variation of the winding pattern in the Wilson loop spectrum when additional trivial bands are added~\cite{ahn2018linking,bouhon2018wilson}.
It has been argued in recent studies~\cite{cano2018topology,wang2018higher,ahn2018linking,bouhon2018wilson} that the fragility of the winding pattern in the Wilson loop spectrum reflects the fragility of the Wannier obstruction.
Here we show concretely that the nontrivial second Stiefel-Whitney class ($w_2=1$) does not induce a Wannier obstruction when the number of bands is bigger than two.
However, this does not mean that an insulator with the nontrivial $w_2$, dubbed a Stiefel-Whitney insulator~\cite{ahn2018linking}, is featureless.
As shown in~\cite{wang2018higher}, anomalous corner states can exist in Stiefel-Whitney insulators, which can be stabilized when additional chiral symmetry is present. We show that the corner charges are induced by the configuration of the Wannier centers constrained by the non-trivial second Stiefel-Whitney class.

\subsection{Reduction of winding numbers from $Z$ to $Z_2$}

Let us first clarify the meaning that the winding number of a vortex reduces from $Z$ to $Z_2$ when the number of bands is increased from two to more than two.
The reduction is due to the ambiguity in the sign of the winding number in the presence of a Dirac string, which was introduced before.
We show that this puts a global constraint on the pair creation and annihilation processes of vortices.

For instance, let us consider a two-band system (band 1 and band 2) with two vortices $v_1$ and $v_2$ with the same winding number.
One can add a trivial band (band 0) below the band minimum of the two-band system. 
When a band inversion happens between band 0 and band 1, two new vortices $v_3$ and $v_4$ with the opposite winding numbers can be created.
Between $v_3$ and $v_4$, a Dirac string exists across which the winding number of $v_1$ or $v_2$ changes its sign.
Then, $v_1$ and $v_2$ can be pair-annihilated after one of them crosses the Dirac string.
If the pair annihilation occurs across the Brillouin zone boundary as shown in Fig.~\ref{fig.fragile}(c,d), $v_3$ and $v_4$ also can be annihilated across the other Brillouin zone boundary. Thus, eventually, each band is decoupled from other bands without any band crossing inbetween. The pair annihilation of $v_1$ and $v_{2}$, which had the same winding number in the absence of band 0, indicates that the integer winding number is not well-defined anymore after the addition of band 0. A quantized Berry phase, which is a $Z_{2}$ invariant, would be the only remaining invariant assigned to each vortex. 

Interestingly, such pair annihilations of two pairs of vortices leave behind two Dirac strings, each encircling a non-contractible cycle in the Brillouin zone.
Let us note that the torus geometry of the Brillouin zone is essential to complete this process.
The appearance of two orthogonal closed Dirac strings indicates that the bands 0, 1, 2 acquire nontrivial Berry phases along the $k_{x}$ and $k_{y}$ cycles, $\Phi_{x}$ and $\Phi_{y}$ such that bands 0, 1, and 2 have $(\Phi_{x},\Phi_{y})=(0,\pi)$, $(\pi,\pi)$, and $(\pi,0)$, respectively, after the completion of the pair annihilation process.

\subsection{Absence of the Wannier obstruction}

As for the Wannier obstruction, let us note that each decoupled band after pair annihilation of vortices is Wannier-representable since a single isolated band with zero Chern number always has a Wannier representation~\cite{alexandradinata2018no}. In fact, the Wannier representation is allowed even if the vortices $v_{1}$ and $v_{2}$ exist after the addition of the trivial band 0.
This is beause the corresponding transition functions can be diagonalized after a suitable gauge transformation, which mixes energy eigenstates at each ${\bf k}$ in general, while keeping the Hamiltonian intact.

Let us note that the Wannier centers for three bands 0, 1, 2 are uniquely determined here.
Since the Berry phases for bands 0, 1, and 2 are $(\Phi_{1},\Phi_{2})=(0,\pi)$, $(\pi,\pi)$, and $(\pi,0)$, respectively,
the relevant Wannier centers are given by $(0,a_2/2)$, $(a_1/2,a_2/2)$, and $(a_1/2,0)$, because $\frac{1}{2\pi}(a_1\Phi_{1},a_2\Phi_{2})$ corresponds to the Wannier center, where $a_{i=1,2}$ are lattice constants.
This can be shown as follows.
Let us recall that the Wannier center of the $n$th band is related to a Berry connection by
\begin{align}
{\bf W}_n
=\braket{n{\bf 0}|\hat{\bf r}|n{\bf 0}}
=V_{\rm cell}\int_{BZ} \frac{d^2k}{(2\pi)^2}{\bf A}_n,
\end{align}
where $V_{\rm cell}$ is the volume of the unit cell, and $\ket{n{\bf R}}$ is the Wannier state of the $n$th band ${\cal B}_n$.
Then, because of the quantization of the Berry phase,
\begin{align}
({\bf W}_{n})_i
&=V_{\rm cell}\int \frac{dk_{\perp}}{2\pi}\left(\int \frac{dk_i}{2\pi}({\bf A}_{n})_i\right)
\nonumber\\
&=V_{\rm cell}\int \frac{dk_{\perp}}{2\pi}\left(\frac{\Phi_{i}({\cal B}_n)}{2\pi}\right)
=\frac{a_i}{2}\frac{\Phi_i({\cal B}_n)}{\pi}.
\end{align}

Let us note that although the three bands have a Wannier representation, the second Stiefel-Whitney class is still nontrivial. This fact can be confirmed by using the Whitney sum formula~\cite{ahn2018linking,hatcher2003vector} for the second Stiefel-Whitney class in the following way.
When all the bands are energetically decoupled, the second Stiefel-Whitney class of the whole bands ${\cal B}\equiv \oplus_n{\cal B}_n$ ($n=0,1,2$) satisfies
\begin{align}
\label{eq.Whitney_sum}
w_{2}({\cal B})
=\frac{1}{\pi^2}\sum_{n\ne m}\Phi_{1}({\cal B}_n)\Phi_2({\cal B}_m)
=4\sum_{n\ne m}({\bf W}_n)_{1}({\bf W}_{m})_{2}.
\end{align}
From the Berry phases for bands 0, 1, 2, given by $(\Phi_{1},\Phi_{2})=(0,\pi)$, $(\pi,\pi)$, and $(\pi,0)$, one can easily find $w_2=1$.

\subsection{Second-order topology characterized by the nontrivial second Stiefel-Whitney class}
\label{subsec.higher_order}

\begin{figure}[t!]
\includegraphics[width=8.5cm]{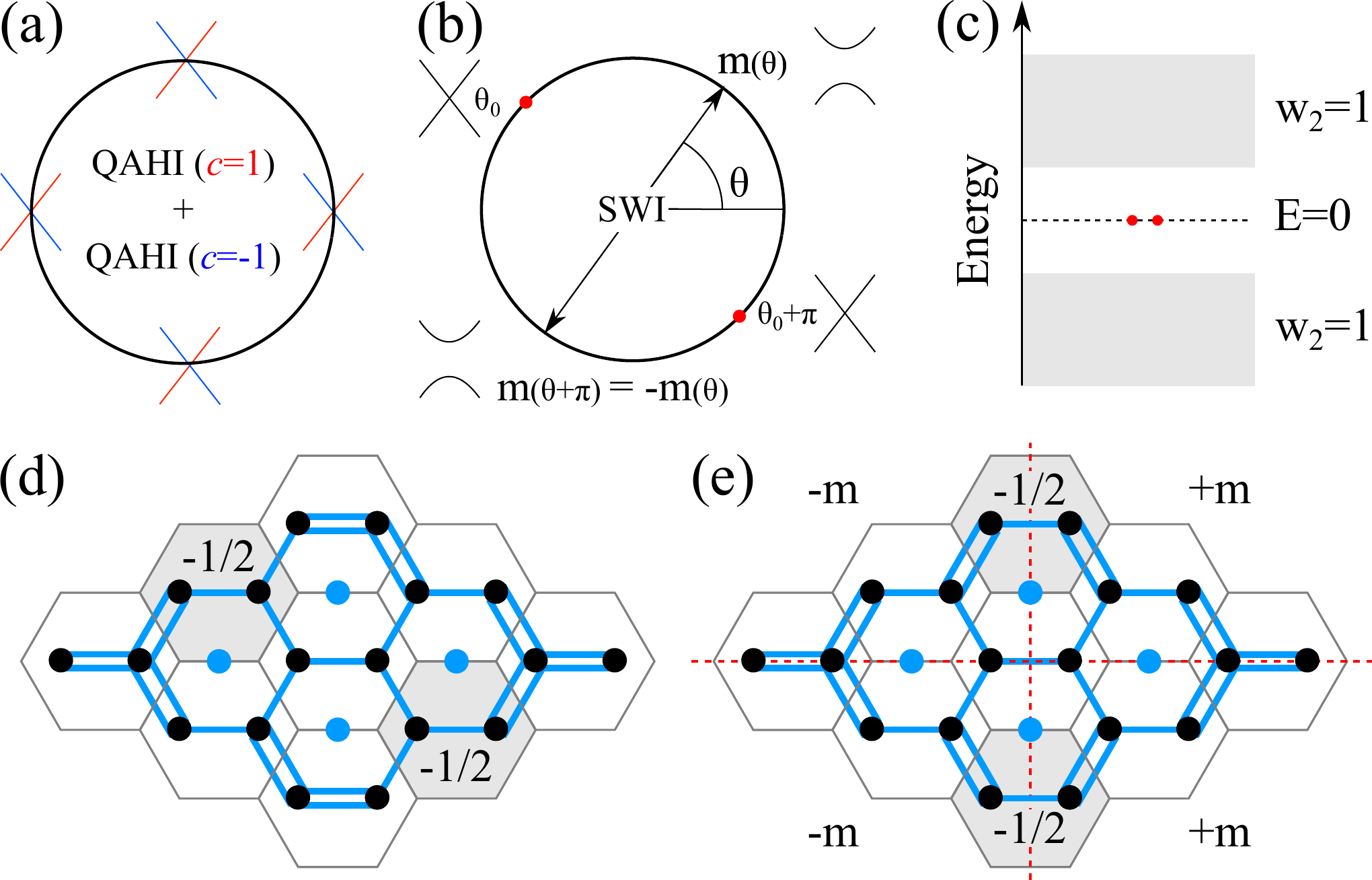}
\caption{Corner states in Stiefel-Whitney insulators.
(a) A system composed of two copies of quantum Hall insulators (QAHI) with two counter-propagating edge states, which can be considered as a particular example of Stiefel-Whitney insulators (SWI).
(b) A Stiefel-Whitney insulator with additional chiral symmetry.
A mass term $m(\theta)$ compatible with $I_{ST}$ and chiral symmetries can open a gap at the edge except at two (mod 4) isolated points.
(c) Finite-size spectrum of the chiral-symmetric Stiefel-Whitney insulator.
(d,e) A schematic figure describing the charge distribution for a finite-size Stiefel-Whitney insulator on the honeycomb lattice without (d) and with (e) a mirror symmetry, respectively.
Mirror symmetry pins corner charges at mirror-invariant points.
Blue dots and links represent localized electric charges. For charge counting, when a dot or a link is shared by two unit cells, we assume that each involved unit cell takes a half of the relevant localized charge.
Here the honeycomb lattice with black dots indicates the finite-size lattice structure whereas the gray honeycomb lattice underneath describes an array of the hexagonal unit cells, each of which contains two black dots in the middle. 
The number $-1/2$ shows the number of localized electrons or the integrated probability density in the unit cell.
}
\label{fig.corner}
\end{figure}

Even though $w_2=1$ states are Wannier-representable, it does not mean that there is no physical consequence associated with them.
Let us note that $w_2$ in Eq.~\eqref{eq.Whitney_sum} has the form of the electric quadrupole moment $q_{xy}$~\cite{benalcazar2017quantized,benalcazar2017electric,franca2018anomalous} as pointed out in~\cite{ahn2018linking}: 
\begin{align}
w_2({\cal B})
&=4\sum_{n\ne m}({\bf W}_n)_{1}({\bf W}_{m})_{2}
\nonumber\\
&=4\sum_{n,m}({\bf W}_n)_{1}({\bf W}_m)_{2}-4\sum_{n}({\bf W}_n)_{1}({\bf W}_{n})_{2},
\nonumber\\
&=4\sum_{n}({\bf W}_n)_{1}({\bf W}_{n})_{2}~~(\text{mod}~~2)\notag\\
&=4q_{xy}~~(\text{mod}~~2),
\end{align}
where we used Eq.~\eqref{eq.Whitney_sum} in the first line and considered trivial total polarization $\sum_{n}({\bf W}_n)_{1}=\sum_{n}({\bf W}_n)_{2}=0$ in the third line.
In fact, anomalous corner states can be induced in systems with $w_2=1$ as shown in~\cite{wang2018higher}.

The presence of corner charges can be understood as follows~\cite{wang2018higher}.
Suppose that a two-dimensional system is composed of two quantum Hall insulators with Chern numbers $c=1$ and $c=-1$, respectively, which are related to each other by $I_{ST}$ [Fig.~\ref{fig.corner}(a)].
This system is a Stiefel-Whitney insulator with $w_2=1$, which can be confirmed by the winding pattern of the Wilson loop spectrum.
For example, the Wilson loop spectrum in Fig.~\ref{fig.TB}(d) is composed of two spectral flows, one going upward and the other going downward, each of which corresponds to $c=1$ and $c=-1$, respectively.
In this particular limit of the Stiefel-Whitney insulator, two counter-propagating chiral edge states exist [Fig.~\ref{fig.corner}(a)].
The edge states are fully gapped after two $I_{ST}$-symmetric mass terms satisfying $m_{1,2}(\theta)=-m_{1,2}(-\theta)$ are added.
Each of the two mass terms has $4N_{i=1,2}+2$ zeros due to the $I_{ST}$ symmetry condition, where $N_{i=1,2}$ are non-negative integers, but the band gap of the edge spectrum $2m=2\sqrt{m_{1}^2+m_2^2}$ is nonzero because $m_1$ and $m_2$ do not vanish simultaneously in general.
However, when there is additional chiral symmetry, only one mass term, which we take it here as $m_1$, remains, so the edge band gap closes at $4N_1+2$ points [See Appendix~\ref{sec.SW-SOTI} for details, where we reproduce the results in~\cite{wang2018higher}].
As the points of zero mass are domain kinks, half charges are localized there.
The corner charges are robust because they are energetically isolated from the bulk bands as shown in Fig.~\ref{fig.corner}(c).
Even when chiral symmetry is broken, the corner charges remain localized as long as they are in the bulk gap.

Alternatively, we can understand the origin of the anomalous corner charges in terms of localized Wannier centers when the number of occupied bands is bigger than two.
For convenience, let us consider a hexagonal lattice with four electrons per unit cell.
Suppose that atoms are located at the corners of the hexagon, and each atom has two electrons.
The Wannier centers for four electrons, which are compatible with the lattice symmetry and the condition $w_2=1$, are then given by $(0,0)$, $(0,a_2/2)$, $(a_1/2,a_2/2)$, and $(a_1/2,0)$, respectively, where $(0,0)$ indicates the center of the hexagon.
Because of the nontrivial Wannier centers associated with the bulk invariant $w_2=1$, an even number of fractional corner charges appear on the edge.
Since the location of corner charges are not constrained by symmetry, they are located at generic positions as shown in Fig.~\ref{fig.corner}(d).
As long as $I_{ST}$ and chiral symmetries are preserved, the half corner charges should appear on the edge since the chiral symmetry requires the corner charges to be in-gap states such that they cannot merge into the bulk state.
Even if extra charges are added $I_{ST}$-symmetrically, the half-integral value of the corner charges is preserved.

\subsection{Role of additional symmetries}
\label{ssec.additional_symmetries}

\begin{figure}[t!]
\includegraphics[width=8.5cm]{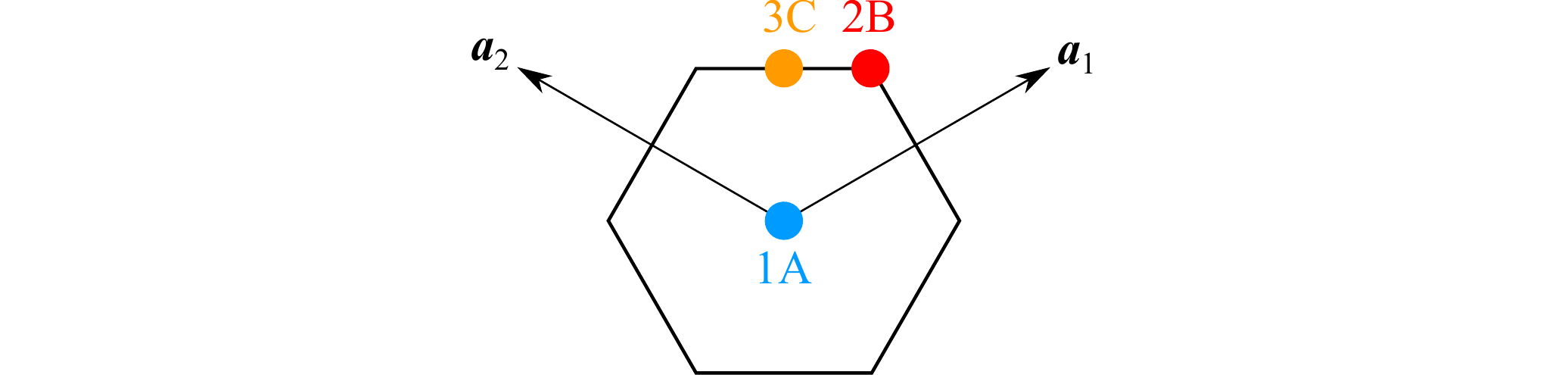}
\caption{
Wyckoff positions in the hexagonal unit cell.
$A=(0,0)$,
$B=\{(\frac{1}{3},\frac{1}{3}),(\frac{2}{3},\frac{2}{3})\}$, and
$C=\{(\frac{1}{2},0),(\frac{1}{2},\frac{1}{2}),(0,\frac{1}{2})\}$
in the basis of lattice vectors ${\bm a}_1$ and ${\bm a}_2$.
}
\label{fig.Wyckoff}
\end{figure}

Up to now, we have focused on the role of $I_{ST}$ symmetry on the band topology.
Let us now discuss the effect of additional $C_{2x}$ and $C_
{3z}$ symmetries that are present in the Moir\'e superlattice of twisted bilayer graphene.

Recently, it has been shown in~\cite{po2018faithful} that the band topology of twisted bilayer graphene remains fragile even in the presence of additional symmetries.
Let us briefly recap the idea of~\cite{po2018faithful} in the context of our theory.
$C_{2x}$ symmetry relates the winding number of the Dirac points at $K$ and $K'$ points as pointed out in ~\cite{po2018origin,zou2018band}.
On the other hand, $C_{3z}$ symmetry pins the vortices at the $K$ and $K'$ points.
Since pair annihilation cannot occur in this case, the Wannier obstruction of the $e_2=1$ phase seems to be stable in the presence of $C_{3z}$ symmetry. This may sound contradictory to the results of~\cite{po2018faithful} in which it was explicitly shown that the Wannier obstruction for nearly flat bands in twisted bilayer graphene disappears after adding $p_x+ip_y$ and $p_x-ip_y$ orbitals at each point of the hexagonal lattice (the Wyckoff position B).  However, in reality, there is no contradiction because pair-annihilation is not required to have Wanner representation, although pair-annihilation of vortices is the clearest way to show the absence of Wannier obstruction when only $I_{ST}$ symmetry is concerned.

In general, the absence of the Wannier obstruction can be proved by showing the existence of a transformation that takes energy eigenstates to other basis states that are Wannier representable.
To illustrate the idea, let us neglect $C_{2x}$ and $C_{3z}$ symmetries for a moment, and focus on the Wannier obstruction related with $I_{ST}$ symmetry. 
Without loss of generality, we can consider the Wannier representation of $I_{ST}$ acting on a set of Wannier states that are individually invariant under $I_{ST}$.
If a Wannier state is not invariant under $I_{ST}$, it has a partner that is related to it by $I_{ST}$.
In this case, by taking a linear combination of the two $I_{ST}$-related states, one can construct a bonding and an anti-bonding states, each of which is invariant under $I_{ST}$ and exponentially localized.
Therefore, when we use $I_{ST}$-symmetric Wannier states as a basis, $I_{ST}$ operator can be represented by a diagonal matrix. 
In momentum space, this indicates that each Bloch state is also not related to other Bloch states under $I_{ST}$ symmetry. 
In this case, according to the Whitney sum formula, the $w_2$ of the whole system is given by the product of $w_1$ for each state in the $I_{ST}$ diagonal basis~\cite{ahn2018higher}. 
Let us note that although $I_{ST}$ operator is diagonal in this basis, the corresponding energy spectrum can be degenerate, possibly with Dirac points between the bands. 
The Whitney sum formula can relate the total $w_2$ to the $w_1$ of Wannier-representable states, as long as the basis diagonalizes $I_{ST}$ operator, independent of the energy spectrum.

Next, let us consider the problem of finding a Wannier representation for three states ${\cal B}$ with the total Berry phase $\Phi_{1}({\cal B})=\Phi_{2}({\cal B})=0$ and the total second Stiefel-Whitney class $w_2({\cal B})=1$.
Since $\Phi_{1}({\cal B})$, $\Phi_{2}({\cal B})$, and $w_2({\cal B})$ are the only topological constraints to gauge transformations, one can gauge transform to three independent $I_{ST}$-symmetric Wannier states with $(\Phi_1,\Phi_2)=(0,\pi)$, $(\pi,\pi)$, and $(\pi,0)$, which have the same total second Stiefel-Whitney class and the total Berry phases.
In this gauge transformation process, the final Wannier-representable states are formed by a linear combination of energy eigenstates, which diagonalize $I_{ST}$ operator. It is not necessary to annihilate vortices during this process.
Thus, the $e_2=1$ phase becomes Wannier-representable after adding one trivial band even without a pair annihilation process.
This can be contrasted to the case of a two-band system with $e_2=1$ and zero total Berry phase where one can never find a $I_{ST}$ diagonal basis compatible with the Whitney sum formula.
This is consistent with the fact that a two-band system with $e_2=1$ and zero total Berry phase is not Wannier-representable.

On the other hand, in the presence of additional $C_{2x}$ and $C_{3z}$ symmetries, the relevant symmetry representation at high symmetry points as well as the topological constraint due to $C_{2z}T$ should be matched simultaneously when we prove the fragile topology of $e_2=1$ phase by adding trivial bands. Formally, the condition for a fragile band topology can be represented as $(e_2=1)\oplus X=Y$ where $X$ and $Y$ indicate sets of Wannier-representable bands. This process has been done in~\cite{po2018faithful}.
Let us briefly review some key ideas in~\cite{po2018faithful}.
Figure~\ref{fig.Wyckoff} shows the Wyckoff positions $A$, $B$, and $C$ in the hexagonal unit cell, and lattice vectors ${\bm a}_1$ and ${\bm a}_2$.
The Wannier orbitals localized at $A=(0,0)$ have $w_2=0$ according to the Whitney sum formula, and those at $B=\{(1/3,1/3),(2/3,2/3)\}$ also have $w_2=0$ because they can be adiabatically moved to $A$ after breaking $C_{3z}$ symmetry while preserving $C_{2z}T$ symmetry.
Only the Wannier orbitals at $C=\{(1/2,0),(1/2,1/2),(0,1/2)\}$ have $w_2=1$.
Thus, the number of Wannier orbitals located at $C$ in the set $Y$ should be different from that in the set X by an odd integer because of the $w_2$ matching condition, $w_2(Y)=w_2(X)+1$.
In addition to this, symmetry representations of $C_{2x}$ and $C_{3z}$ should be matched at the relevant high-symmetry points in the momentum space.
While it is possible to take $X$ as a single band orbital in the absence of $C_{2x}$ and $C_{3z}$ symmetries, the symmetry representation matching can increase the minimum number of bands in $X$.
As shown in~\cite{po2018faithful}, when the $(e_2=1)$ phase is the nearly flat bands of twisted bilayer graphene, the minimal $X$ consists of three orbitals, which can be chosen to be $p_z$ (or $s$) orbitals at the Wyckoff position $C$. Then, the resulting $Y$ is composed of $s$ and $p_\pm$ (or $p_z$ and $p_\pm$) orbitals at the Wyckoff position $A$ in addition to $p_z$ (or $s$) orbitals at the Wyckoff position $B$. For completeness, we discuss the details of this procedure in Appendix~\ref{sec.fragility_symmetries}.

The additional symmetries also constrain the location of anomalous corner charges.
While $C_{3z}$ symmetry just requires    tthat corner charges appear $C_{3z}$-symmetrically, $C_{2x}$ symmetry puts a stronger constraint that corner charges are located at either $C_{2x}$- or $C_{2x}I_{ST}$-invariant corners but not at both.
Notice that, when both $C_{2x}$ and $I_{ST}=C_{2z}T$ are symmetry operators, $C_{2y}T=C_{2x}I_{ST}:(x,y)\rightarrow (-x,y)$ is also a symmetry operator.
We thus have two effective mirror symmetries under $\tilde{M}_x\equiv C_{2y}T$ and $\tilde{M}_y\equiv C_{2x}$.
As shown in Fig.~\ref{fig.corner}(b), one mirror operation changes the sign of the edge mass whereas the other does not, because the product $\tilde{M}_x\tilde{M}_y=I_{ST}$ changes the sign of the edge mass [See Appendix~\ref{sec.SW-SOTI} for more details].
Consequently, anomalous charges will appear at either $C_{2x}$- or $C_{2x}I_{ST}$-invariant corners but not at both.

\begin{figure}[t!]
\includegraphics[width=8.5cm]{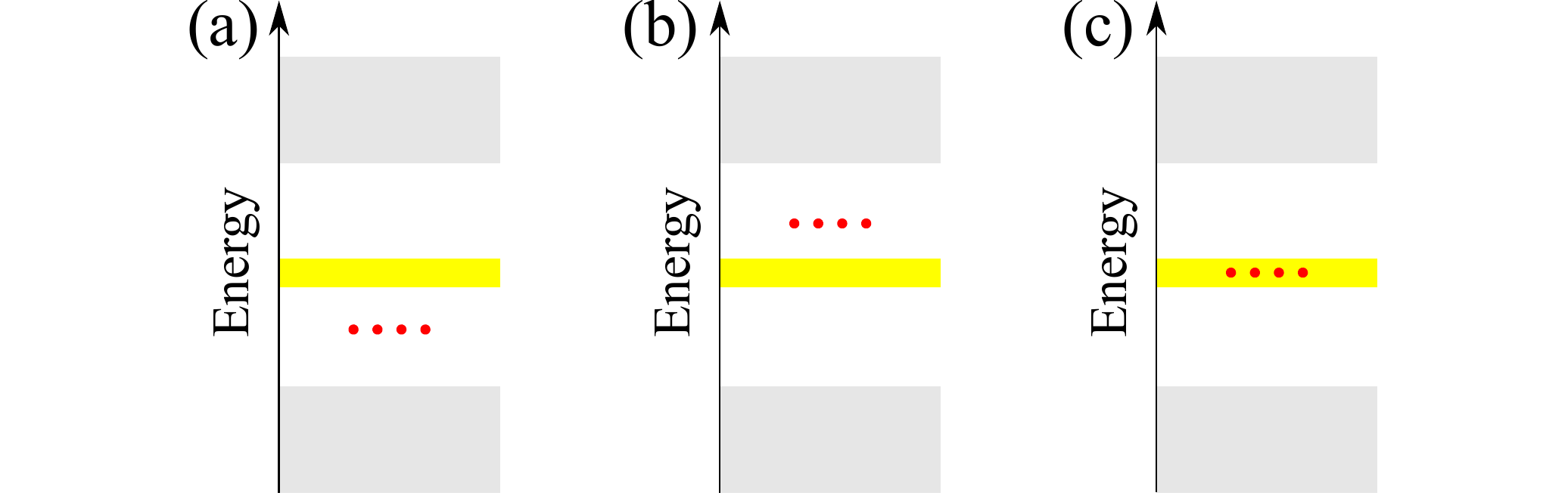}
\caption{
Possible energy spectra which may support corner charges in twisted bilayer graphene.
Shaded grey regions correspond to the bulk bands, where the middle yellow band corresponds to the nearly flat bands near the charge neutrality.
Corner charges may be (a,b) in the bulk gap or (c) merged with the middle band.
As each of the middle bands arising from the two decoupled valley degrees of freedom has $e_2=1$, the corner charge is doubled as compared to Fig.~\ref{fig.corner}. 
When spin degrees of freedom are taken into account, the amount of corner charges should also be doubled further.
}
\label{fig.TBG}
\end{figure}

Let us finally discuss the possibility that anomalous corner charges appear in twisted bilayer graphene. Fig.~\ref{fig.TBG} (a) and (b) show the two possible energy spectra which can support localized corner states. As we have explained in Sec.~\ref{subsec.higher_order}, bands with $w_2=1$ can induce corner charges. In the case of TBG, two bands arising from any one of the valleys has $e_2=1$ in the absence of intervalley coupling. Therefore, by tuning the Fermi level, corner charges can be induced. Because the two valleys are time-reversal partners, we expect that the corner charges derived from two valleys, will have the same energy. On the other hand, in the presence of the approximate chiral symmetry, the situation shown in Fig.~\ref{fig.TBG}(c) would occur. Then, the localized corner charge cannot exist due to the hybridization with the bulk bands. To confirm the presence of corner charges, more sophisticated first-principles band structure calculation should be performed.

\section{Conclusion}
\label{sec.conclusion}

We have shown that the Euler class of real two band systems with $I_{ST}$ symmetry is identical to the total winding number of the band degeneracies between two bands.
Namely, the topological charge of band crossing points determines the global band topology.
We expect that our theory here can be generalized to a broader class of systems.
Recently, a no-go theorem was proposed in~\cite{alexandradinata2018no}: the statement is that Wannier obstruction of a single band can come only from a nontrivial first Chern number. 
This implies that Wannier obstructions originating from the other topological invariants describing multi-band systems may require unremovable band degeneracies.
It would be an interesting topic for future studies to establish the general relationship between the symmetry eigenvalues at high symmetry points, the topological charge of band degeneracies and the global band topology in crystalline topological materials.

\begin{acknowledgments}
We acknowledge the helpful discussions with  Aris Alexandradianata, Yoonseok Hwang, Benjamin J. Wieder, and Bogdan Andrei Bernevig.
J.A. and S.P. were supported by IBS-R009-D1.
B.-J.Y. was supported by the Institute for Basic Science in Korea (Grant No. IBS-R009-D1) and Basic Science Research Program through the National Research Foundation of Korea (NRF) (Grant No. 0426-20170012, No.0426-20180011), and  the POSCO Science Fellowship of POSCO TJ Park Foundation (No.0426-20180002).
This work was supported in part by the U.S. Army Research Office under Grant Number W911NF-18-1-0137.
We also acknowledge Haruki Watanabe for hosting a wonderful workshop ``Symmetry and Topology in Condensed-Matter Physics" during which we got the initial idea for this work.
\end{acknowledgments}

{\it Note added.|} During the preparation of our manuscript, we have found  related works~\cite{song2018all,po2018faithful}. In \cite{song2018all}, based on first-principles calculations, it was found that the Wilson loop spectrum for nearly flat bands in twisted bilayer graphene has nontrivial winding, which is consistent with our conclusion.
In~\cite{po2018faithful}, tight-binding models were constructed explicitly by adding trivial bands to the two nearly flat bands, which demonstrates the fragility of the Wannier obstruction for the flat bands.

After the completion our manuscript, we became aware of related mathematical studies~\cite{mathai2017differential,mathai2017global,thiang2017fu} in which the relationship between the topological charges of Dirac and Weyl points and the global topology is examined in more abstract settings. 
In particular, it was also pointed out in~\cite{mathai2017differential} that the total winding number of vortices is given by the Euler class, based on the generalized Poincar\'e-Hopf theorem.

Just after our first manuscript was uploaded on arXiv, an independent study~\cite{wu2018beyond} appeared on arXiv, where non-abelian topological properties of nodal lines in $PT$-symmetric 3D spinless fermion systems are systematically studied by using homotopy theory.
Applying the idea proposed in~\cite{wu2018beyond} to 2D problems, one can find that the sign reversal of vortices across a Dirac string proposed in our paper can be interpreted as a manifestation of the non-abelian algebra of topological charges.

\appendix

\section{Winding number in general chiral symmetric systems}
\label{sec.chiral}

In the main text, we showed that the winding number can be computed using the off-diagonal Berry phase for $I_{ST}$-symmetric two bands.
Notice that $I_{ST}$-symmetric two band Hamiltonian have chiral symmetry when the chemical potential term, which is irrelevant for the band crossing, is neglected.
Here we show that the same method can be applied to any chiral symmetric systems.

\subsection{Sewing matrix and the Berry phase}

Consider the sewing matrix for chiral symmetry operator $S$:
\begin{align}
S_{mn}({\bf k})\equiv \braket{u_{m\bf k}|S|u_{n\bf k}}.
\end{align}
It takes an off-diagonal form
\begin{align}
S({\bf k})
=
\begin{pmatrix}
0&s^{-1}({\bf k})\\
s({\bf k})&0
\end{pmatrix}
\end{align}
in the basis
\begin{align}
\ket{u_{\bf k}}
=
\begin{pmatrix}
\ket{u^{\text{unocc}}_{\bf k}}\\
\ket{u^{\text{occ}}_{\bf k}}
\end{pmatrix},
\end{align}
where $s({\bf k})\in U(N)$.
The Berry connection
\begin{align}
A_{mn}({\bf k})=\braket{u_{m\bf k}|\nabla_{\bf k}|u_{n\bf k}}
\end{align}
in chiral symmetric systems satisfies
\begin{align}
{\bf A}({\bf k})=S^{-1}({\bf k}){\bf A}({\bf k})S({\bf k})+S^{-1}({\bf k})\nabla_{\bf k}S({\bf k}),
\end{align}
which shows that
\begin{align}
{\bf A}_2({\bf k})
&=s{\bf A}_1({\bf k})s^{-1}+s({\bf k})\nabla_{\bf k}s^{-1}({\bf k}),
\end{align}
and
\begin{align}
{\bf a}({\bf k})\equiv
i{\bf A}_{12}({\bf k})s({\bf k})
=\left(i{\bf A}_{12}({\bf k})s({\bf k})\right)^{\dagger}
={\bf a}^{\dagger}({\bf k}).
\end{align}
Accordingly, the Berry connection takes the following form.
\begin{align}
{\bf A}({\bf k})
&=
\begin{pmatrix}
{\bf A}_1({\bf k})
&-i{\bf a}({\bf k})s^{-1}({\bf k})\\
-is({\bf k}){\bf a}({\bf k})
&s{\bf A}_1({\bf k})s^{-1}+s({\bf k})\nabla_{\bf k}s^{-1}({\bf k})
\end{pmatrix},
\end{align}
where ${\bf A}_1({\bf k})$ and ${\bf a}({\bf k})$ are undetermined by the sewing matrix for chiral symmetry.

Under a gauge transformation $\ket{u_{n\bf k}}\rightarrow \ket{u'_{n\bf k}}=U_{mn}({\bf k})\ket{u_{m\bf k}}$, the sewing matrix transforms as
\begin{align}
S({\bf k})\rightarrow S'({\bf k})=U^{\dagger}({\bf k})S({\bf k})U({\bf k}).
\end{align}
Accordingly, under a diagonal gauge transformation
\begin{align}
U({\bf k})
&=
\begin{pmatrix}
U_1({\bf k})&0\\
0&U_2({\bf k})
\end{pmatrix},
\end{align}
\begin{align}
s^{-1}({\bf k})\rightarrow s'^{-1}({\bf k})=U_1^{\dagger}({\bf k})s^{-1}({\bf k})U_2({\bf k}).
\end{align}
Since the Berry connection transforms by
\begin{align}
{\bf A}_{12}({\bf k})
\rightarrow 
{\bf A}'_{12}({\bf k})
&=U_1^{\dagger}({\bf k}){\bf A}_{12}({\bf k})U_2({\bf k}),
\end{align}
we get
\begin{align}
{\bf a}({\bf k})
\rightarrow 
{\bf a}'({\bf k})
&=U_1^{\dagger}({\bf k}){\bf a}({\bf k})U_1({\bf k}).
\end{align}

Notice that the matrix trace of any power of ${\bf a}({\bf k})$ is gauge invariant.
It suggests that $\oint_{S^d}{\rm Tr}[{\bf a}^d({\bf k})]$ may serve as a $d$-dimensional topological invariant.

\subsection{Winding number and the off-diagonal Berry phase}

Suppose that the unoccupied and occupied bands are topologically trivial as a whole.
Then there exists a Hamiltonian which is smooth over the whole Brillouin zone that describes both the unoccupied and occupied bands.
When the chiral operator is represented by
\begin{align}
S=
\begin{pmatrix}
1_{\rm N\times N}&0\\
0&-1_{\rm N\times N}
\end{pmatrix},
\end{align}
the chiral-symmetric Hamiltonian takes the form of
\begin{align}
H({\bf k})
&=
\begin{pmatrix}
0&h({\bf k})\\
h^{\dagger}({\bf k})&0
\end{pmatrix}\notag\\
&=
\begin{pmatrix}
0&U({\bf k})P({\bf k})\\
P({\bf k})U^{\dagger}({\bf k})&0
\end{pmatrix},
\end{align}
where we used polar decomposition of $h({\bf k})$, where $U({\bf k})\in U(N)$ and $P({\bf k})=\sqrt{h^{\dagger}({\bf k})h({\bf k})}$.
The energy eigenstates on gapped regions are given by
\begin{align}
\ket{u_{n\bf k}^{\text{unocc}}}
&=\frac{1}{\sqrt{2}}
\begin{pmatrix}
U({\bf k})\ket{e_{n\bf k}}\\
\ket{e_{n\bf k}}
\end{pmatrix},\notag\\
\ket{u_{n\bf k}^{\text{occ}}}
&=\frac{1}{\sqrt{2}}
\begin{pmatrix}
U({\bf k})\ket{e_{n\bf k}}\\
-\ket{e_{n\bf k}}
\end{pmatrix},
\end{align}
where $\ket{e_{n=1,...,N,\bf k}}$ are the eigenstates of $P({\bf k})$ with eigenvalues $|E_{n\bf k}|$, and $\ket{u_{n\bf k}^{\text{unocc}}}$ and $\ket{u_{n\bf k}^{\text{occ}}}$ have energies $|E_{n\bf k}|$ and $-|E_{n\bf k}|$, respectively.

In this choice of gauge, the Berry connection is given by
\begin{align}
{\bf A}({\bf k})
=&
\begin{pmatrix}
\frac{1}{2}U^{\dagger}({\bf k})\nabla_{\bf k}U({\bf k})
&(-i)\frac{i}{2}U^{\dagger}({\bf k})\nabla_{\bf k}U({\bf k})\\
(-i)\frac{i}{2}U^{\dagger}({\bf k})\nabla_{\bf k}U({\bf k})
&\frac{1}{2}U^{\dagger}({\bf k})\nabla_{\bf k}U({\bf k})
\end{pmatrix}\notag\\
&+
\begin{pmatrix}
\braket{e_{\bf k}|\nabla_{\bf k}|e_{\bf k}}
&0\\
0
&\braket{e_{\bf k}|\nabla_{\bf k}|e_{\bf k}}
\end{pmatrix}.
\end{align}
From this expression, we see that
\begin{align}
{\rm Tr}[{\bf a}({\bf k})]
=\frac{i}{2}{\rm Tr}[U^{\dagger}({\bf k})\nabla_{\bf k}U({\bf k})]
=\frac{i}{2}\nabla_{\bf k}\log \det U({\bf k}),
\end{align}
such that
\begin{align}
\oint_{S^1} d{\bf k}\cdot {\rm Tr}[{\bf a}({\bf k})]=\frac{1}{2}\oint_{S^1} d{\bf k}\cdot \nabla_{\bf k}\theta({\bf k})=N_w\pi,
\end{align}
where $\det U({\bf k})=\exp(-i\theta({\bf k}))$.
This off-diagonal Berry phase is invariant under gauge transformations which do not mix the unoccupied and occupied bands.
As in the main text, the sign of the winding number is fixed after we choose the global sign of Tr$[{\bf a}({\bf k})]$.

In general, the winding number in $(2n+1)$-dimensional chiral symmetric systems is given by
\begin{align}
N_w^{(2n+1)}
=\frac{(-1)^nn!}{(2n+1)!\pi^{n+1}}\oint_{S^{2n+1}}d^{2n+1}k{\rm Tr}[({\bf a}({\bf k}))^{2n+1}].
\end{align}

\subsection{Space-time symmetries}

Let us investigate the space-time symmetry constraint on ${\bf a}({\bf k})$.
It will turn out that ${\bf a}({\bf k})$ transforms like a Berry curvature rather than a Berry connection.
We get the same conclusion for $I_{ST}$-symmetric two bands because they have effective chiral symmetry if we neglect the chemical potential.

First we consider a crystalline symmetry operator $G$, where
\begin{align}
G_{mn}({\bf k})\equiv \braket{u_{mG\bf k}|G|u_{n\bf k}}.
\end{align}
It takes the form
\begin{align}
G({\bf k})
=
\begin{pmatrix}
g_1({\bf k})&0\\
0&g_2({\bf k})
\end{pmatrix}
\end{align}
in the basis
\begin{align}
\ket{u_{\bf k}}
=
\begin{pmatrix}
\ket{u^{\text{unocc}}_{\bf k}}\\
\ket{u^{\text{occ}}_{\bf k}}
\end{pmatrix},
\end{align}
where $g_{1,2}({\bf k})\in U(N)$.
The symmetry constraint for the Berry connection is given by
\begin{align}
{\bf A}({\bf k})=G^{-1}({\bf k})P_G^{-1}\cdot {\bf A}(P_G{\bf k})G({\bf k})+G^{-1}({\bf k})\nabla_{\bf k}G({\bf k}),
\end{align}
where $P_G^{-1}$ is the point group part of $G$, and $P_G^{-1}\cdot {\bf A}$ indicates the transformation of vector components of ${\bf A}$ under the action of $G$.
Then,
\begin{align}
{\bf A}_{i}({\bf k})=g^{-1}_{i}({\bf k})P_G^{-1}\cdot {\bf A}_{i}(P_G{\bf k})g_{i}({\bf k})+g^{-1}_{i}({\bf k})\nabla_{\bf k}g_{i}({\bf k}),
\end{align}
where $A_1\equiv A_{11}$ and $A_2\equiv A_{22}$, and
\begin{align}
{\bf A}_{12}({\bf k})=g_1^{-1}({\bf k})P_G^{-1}\cdot {\bf A}_{12}(P_G{\bf k})g_2({\bf k}).
\end{align}
Because $[S,G]=0$ requires that
\begin{align}
G({\bf k})S({\bf k})=S({G\bf k})G({\bf k}),
\end{align}
so that 
\begin{align}
s^{-1}({\bf k})=g_1^{-1}({\bf k})s^{-1}(P_G{\bf k})g_2({\bf k}),
\end{align}
we get
\begin{align}
{\bf a}({\bf k})
&=g_1^{-1}({\bf k})P_G^{-1}\cdot {\bf a}(P_G{\bf k})g_1({\bf k})\notag\\
&=g_2^{-1}({\bf k})P_G^{-1}\cdot {\bf a}(P_G{\bf k})g_2({\bf k}).
\end{align}

Next we consider the time-reversal symmetry operator $T$, where
\begin{align}
B_{mn}({\bf k})\equiv \braket{u_{m-\bf k}|T|u_{n\bf k}}.
\end{align}
It takes the form
\begin{align}
B({\bf k})
=
\begin{pmatrix}
b_1({\bf k})&0\\
0&b_2({\bf k})
\end{pmatrix}
\end{align}
in the basis
\begin{align}
\ket{u_{\bf k}}
=
\begin{pmatrix}
\ket{u^{\text{unocc}}_{\bf k}}\\
\ket{u^{\text{occ}}_{\bf k}}
\end{pmatrix},
\end{align}
where $b_{1,2}({\bf k})\in U(N)$.
The symmetry constraint for the Berry connection is given by
\begin{align}
{\bf A}(-{\bf k})=-B({\bf k}){\bf A}^*({\bf k})B^{-1}({\bf k})-B({\bf k})\nabla_{\bf k}B^{-1}({\bf k}).
\end{align}
Then,
\begin{align}
{\bf A}_{i}(-{\bf k})=-b_{i}({\bf k}){\bf A}^{*}_{i}({\bf k})b^{-1}_{i}({\bf k})-b_{i}({\bf k})\nabla_{\bf k}b^{-1}_{i}({\bf k}),
\end{align}
where $A_1\equiv A_{11}$ and $A_2\equiv A_{22}$, and
\begin{align}
{\bf A}_{12}(-{\bf k})=-b_1({\bf k}){\bf A}^*_{12}({\bf k})b_2^{-1}({\bf k}).
\end{align}
Because $[S,T]=0$ requires that
\begin{align}
B({\bf k})S^*({\bf k})=S({-\bf k})B({\bf k}),
\end{align}
so that 
\begin{align}
s^{-1}(-{\bf k})=b_1({\bf k})(s^{-1})^*({\bf k})b_2^{-1}({\bf k}),
\end{align}
we get
\begin{align}
{\bf a}(-{\bf k})
&=b_1({\bf k}){\bf a}^*({\bf k})b_1^{-1}({\bf k})\notag\\
&=b_2({\bf k}){\bf a}^*({\bf k})b_2^{-1}({\bf k}).
\end{align}

Let us finally comment that $PT=K$ and $S=\sigma_y$ anticommutes in the $PT$-symmetric two-band model whereas we have assumed that $S$ commutes with space-time symmetries in this subsection.

\section{Pair annihilation of vortices in terms of monopole nodal lines}
\label{sec.monopole}

\begin{figure}[t!]
\includegraphics[width=8.5cm]{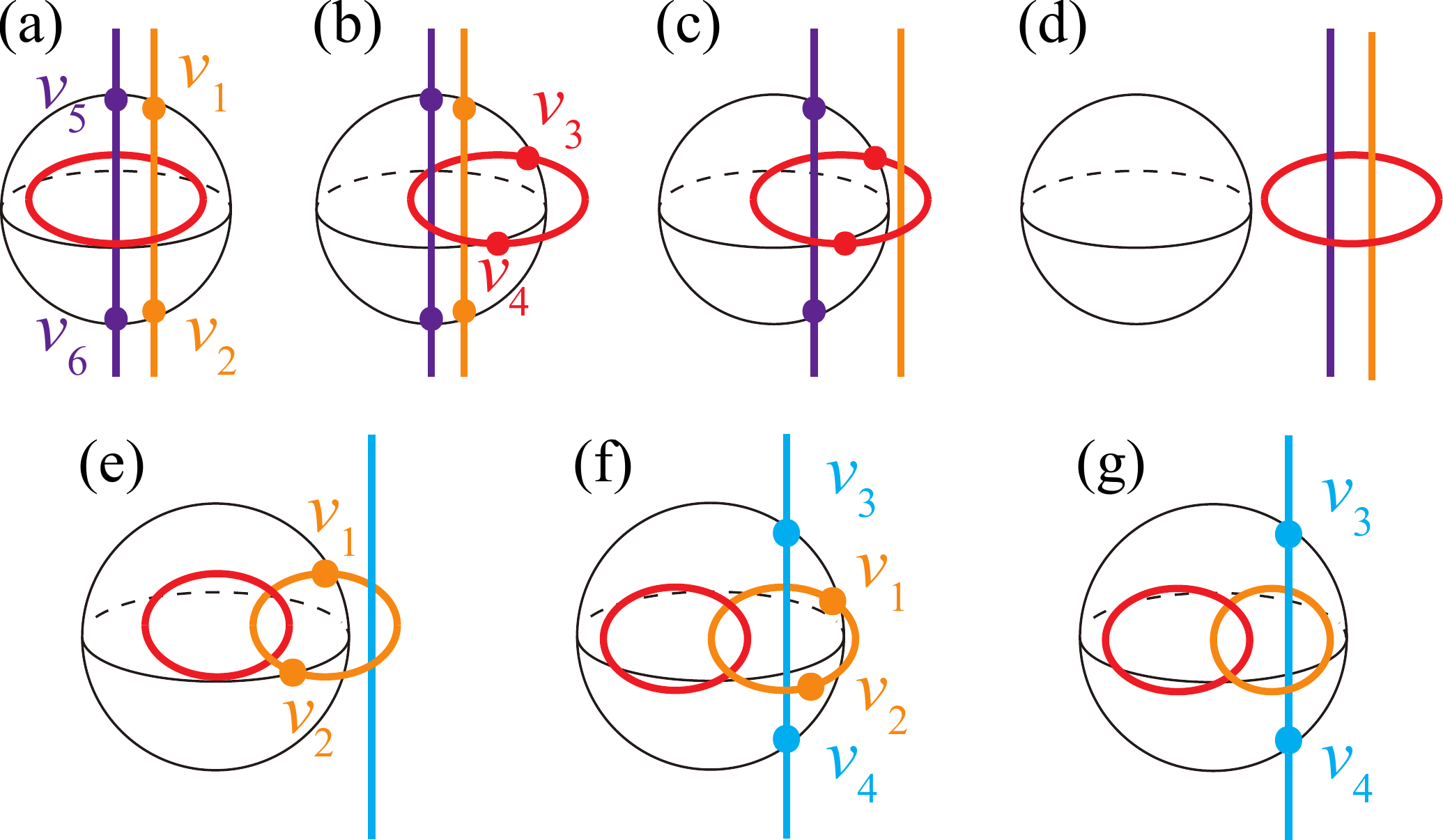}
\caption{Description of the vortex-antivortex pair creation and annihilation in the view of nodal lines.
(a-d) The same process in the point of view of monopole lines in three dimensions (3D). The orange (purple) line is the band crossing points between occupied (unoccupied) bands. The red line is the band crossing points between bands 2 and 3. The sphere can be thought of as the Brillouin zone in (a-e) in Fig.~\ref{fig.tpt1}.
(e-g) Stability of vortices on a sphere against adding a trivial band.
The blue, orange, and red lines are formed between band 0 and band 1, between band 1 and band 2, and between band 2 and band 3, respectively.
(e) Vortices $v_1$ and $v_2$ with the same winding number.
(f) Pair creation of $v_3$ and $v_4$. 
$v_1$ and $v_2$ can be pair-annihilated after crossing the Dirac string existing between vortices $v_3$ and $v_4$.
(g) After the pair annihilation of $v_1$ and $v_2$.
The resulting $v_3$ and $v_4$ have the same winding number.
}
\label{fig.line}
\end{figure}

In the main text, we explained the pair annihilation process of vortices by introducing the Dirac string.
It is possible to give this pair annihilation process an alternative description in terms of a nodal line with a monopole charge in 3D.

\subsection{Topological phase transition}

Let us consider a four-band system with two occupied bands (bands 1, 2) and two unoccupied bands (bands 3, 4), which are now in 3D space. Moreover, let us suppose that the band crossing between bands 2 and 3 forms a monopole nodal line at $E_{F}$. One immediate physical consequence arising from the monopole charge of the nodal line is that another nodal line formed between bands 1 and 2 should be linked with the monopole nodal line as shown in~\cite{ahn2018linking}.
Because of this linking structure, a sphere wrapping the monopole nodal line should cross the other nodal line below $E_{F}$ at two points as shown in Fig.~\ref{fig.line}(a-d).
Considering the wrapping sphere as a 2D BZ, the crossing between the sphere and the nodal line below $E_{F}$ indicates the Dirac points formed between two occupied bands.
Since the monopole charge is identical to the Euler class when the number of occupied bands is two~\cite{ahn2018linking}, the wrapping sphere exactly corresponds to a 2D insulator with $e_{2}=1$ having two vortices with the same winding number between the two occupied bands.  
Note that in Fig.~\ref{fig.line}(a-d), we have also drawn a purple line next to the orange line, to indicate that the gap closing points formed by the unoccupied bands (bands 3 and 4), for which the same comments apply as those for the occupied bands.  Also, the points at which the orange (purple) line crosses the sphere corresponds to the vortices between bands 1 and 2 (bands 3 and 4) in the 2D insulator. Then, we see that for the occupied bands to become trivial, the orange line and the purple line should leave the sphere before the red loop does. The trajectories of the crossing points between the sphere and the three nodal lines (orange, red, purple) correspond to the process shown in Fig.~\ref{fig.tpt1}(a-d).

\subsection{Stability of vortices on sphere}

In Sec.~\ref{sec.fragile} we have shown that the $e_2=1$ phase is fragile because the presence of an additional trivial band allows the pair annihilation of vortices.
Here, we show that the pair annihilation process cannot occur if the Brillouin zone has a spherical geometry. Since all loops are contractible on a sphere, Berry phase should always be trivial.
On a sphere, a pair annihilation of vortices with the same winding number formed between bands 1 and 2 necessarily lead to a pair creation of other vortices with the same winding number between another pair of bands, (for instance, between bands 0 and 1) as shown in Fig.~\ref{fig.tpt1} (a-d).
This is related to the stable linking structure of monopole nodal lines in the 3D Brillouin zone as illustrated in Fig.~\ref{fig.line}(e-g).

\section{Topological phase transition in the presence of spin-orbit coupling}
\label{sec.tpt_soc}

\begin{figure}[t!]
\includegraphics[width=8.5cm]{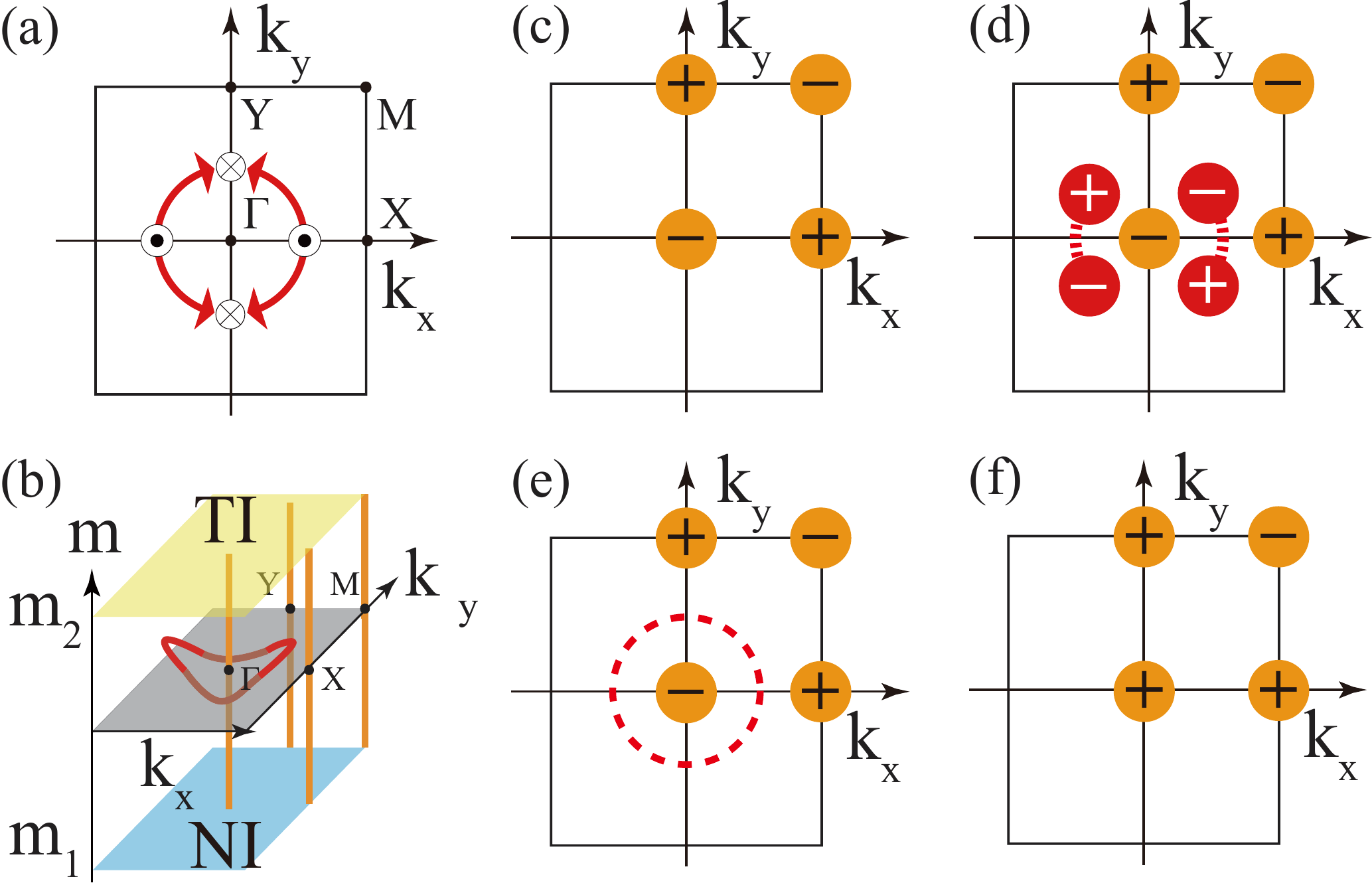}
\caption{Topological phase transition between a normal insulator (NI) and a quantum spin Hall insulator (TI) in spin-orbit coupled noncentrosymmetric systems with $C_{2z}$ and $T$ symmetries.
(a) Trajectory of Dirac points (or vortices) in the intermediate gapless phase.
(b) A nodal line, representing the trajectory of Dirac points in the 3D space $(k_x,k_y,m)$ where $m$ denotes a tuning parameter.
The nodal line (red) encircles the Kramers degenerate lines (orange) an odd number of times.
(c-f) Change of the winding number of the Kramers degenerate points above and below
the Fermi energy $E_{F}$. 
The vortices of Dirac points at $E_{F}$ and Kramers degenerate points are shown in red and orange, respectively.
The $\pm$ sign shows the local winding number of vortices.
Dashed lines in (d) and (e) are Dirac strings across which the winding number of a Kramers degenerate point changes its sign.
The total winding number of Kramers degenerate points below $E_{F}$ (and also above $E_{F}$) changes from zero in (c) to two in (f).
}
\label{fig.tpt2}
\end{figure}
Up to now, we have focused on the case without spin orbit coupling.
Even in the presence of spin-orbit coupling, however, $I_{ST}=C_{2z}T$ acts like a complex conjugation satisfying $(C_{2z}T)^{2}=+1$, so it can protect vortices in the absence of inversion $P$ symmetry.
In a recent work~\cite{ahn2017unconventional}, it was shown that pair creation and pair annihilation of Dirac points can mediate a topological phase transition between a normal insulator and a quantum spin Hall insulator in spin-orbit coupled noncentrosymmetric systems with $T$ and $C_{2z}$ symmetries.
In the course of a topological phase transition, the trajectory of Dirac points form a closed loop surrounding time-reversal-invariant momenta (TRIM) an odd number of times as shown in Fig.~\ref{fig.tpt2}(a). This pair creation and pair annihilation processes can also be understood in terms of the winding number changes across a Dirac string as described below.
Such an alternative description is possible due to the equivalence between the second Stiefel-Whitney class and the Fu-Kane-Mele invariant in the system as explained in detail in Appendix~\ref{sec.SW=FKM}.

\subsection{Topological phase transition}

For convenience, let us suppose that the system is composed of two occupied bands and two unoccupied bands, although the topological phase transition is well-defined when the number of bands is larger.
Because of time reversal symmetry, occupied bands are always degenerate at TRIM.
Then in the 3D space $(k_x,k_y,m)$ including a tuning parameter $m$ controlling the phase transition, the Kramers degeneracies form four straight nodal lines along the $m$-direction as shown in Fig.~\ref{fig.tpt2} (b).
If a normal insulator with $e_2=0$ and a topological insulator with $e_2=1$ exist for $m<m_1$ and $m=m_2>m_1$, respectively, the trajectory of Dirac points corresponds to the intersection between the red nodal line and the constant $m$ planes in Fig.~\ref{fig.tpt2} (b) as $m$ is tuned in the range $m_1<m<m_2$.
Due to the straight nodal lines from Kramers degeneracies, any nodal loop, representing the trajectory of Dirac points, centered at a TRIM should be a monopole line due to the linking structure~\cite{ahn2018linking}.
Then, the shape of the trajectory of Dirac points reflects the correspondence between the closed trajectory of vortices and the relevant change of the topological invariant.

Explicitly, let us explain how the winding number transition is related to the closed trajectory of gap-closing points.
Consider a transition from a normal insulator with $e_2=0$ to a quantum spin Hall insulator with $e_2=1$, and assume, for simplicity, that the band structure of the normal insulator has no degeneracy other than the Kramers degeneracy.
Then, $e_2=0$ indicates that the total winding number of the Kramers degenerate points below the Fermi energy $E_{F}$ should be zero as shown in Fig.~\ref{fig.tpt2}(c).
When the band gap closes and vortices are pair-created at the momentum $\bm{k}$ and $-\bm{k}$, Dirac strings connecting each pair of vortices are also generated [Fig.\ref{fig.tpt2}(d)].
Let us note that both $C_{2z}$ and $T$ require that the winding number of vortices at ${\bf k}$ and $-{\bf k}$ is equal.
As the Dirac strings follow the trajectory of the vortices, they eventually form a closed loop around a TRIM after the pair annihilation of vortices [Fig.~\ref{fig.tpt2}(e)].
Then, the Dirac string can be removed after flipping the sign of the winding number of the Kramers degenerate point encircling [Fig.~\ref{fig.tpt2}(f)].
Because of the sign change, the total winding number of Kramers degenerate points becomes two.
This indicates the change of the topological invariant $e_2$ from zero to one.

\subsection{Equivalence of the second Stiefel-Whitney class and the Fu-Kane-Mele invariant}
\label{sec.SW=FKM}

The Wilson loop method implies that the second Stiefel-Whitney class is identical to the $Z_2$ topological variant, because they are characterized by the same pattern of the Wilson loop spectral flow.
Here we provide another proof of the equivalence using the Euler class and the Fu-Kane-Mele invariant.
Our proof here goes parallel with the derivation of the relation between the second Stiefel-Whitney class and inversion eigenvalues, presented in Supplemental Material of~\cite{ahn2018linking}.

We first notice that the total Berry phase of the occupied bands is always nontrivial in this system because the Berry phase is quantized to be a multiple of $\pi$ due to the $C_{2z}T$ symmetry, but $T$ further requires it be a multiple of $2\pi$ because energy bands form Kramers pairs~\cite{ahn2017unconventional}.
Therefore, the occupied states are always orientable in a real gauge, so we take transition functions belonging to the special orthogonal group.
Furthermore, we will only consider two occupied bands because we can block-diagonalize the sewing matrix $B$ into $2\times 2$ blocks by lifting the accidental degeneracy of occupied bands without loss of generality.

Let us take a real gauge: $C_{2z}T\ket{\tilde{u}_{n\bf k}}=\ket{\tilde{u}_{n\bf k}}$.
Time reversal symmetry imposes a further constraint on energy eigenstates by
\begin{align}
T\ket{u^B_{n\bf k}}=B^{AB}_{mn}({\bf k})\ket{u^A_{m\bf -k}},
\end{align}
where $B({\bf k})\in O(2)$ is the sewing matrix for time reversal, and $A$ and $B$ denotes the local patch on which the states are smoothly defined.

In fact, the sewing matrix $B$ belongs to $SO(2)$.
In general, as the real occupied states are not smooth over the whole 2D Brillouin zone, the sewing matrix also is not smooth.
The sewing matrix defined on $C$ and $D$ patches are related to the one defined on $A$ and $B$ patches as
\begin{align}
B^{CD}({\bf k})=(t^{AC}({\bf -k}))^{-1}B^{AB}({\bf k})t^{BD}({\bf k}),
\end{align}
where $A$ and $C$ covers $-\bf k$, and $B$ and $D$ covers $\bf k$, and $t^{AB}$ and $t^{CD}$ are the transition functions defined by
$\ket{u^C_{n\bf -k}}= t^{AC}_{mn}({\bf  -k})\ket{u^A_{m\bf - k}}$ and $\ket{u^D_{n\bf k}}= t^{BD}_{mn}({\bf k})\ket{u^B_{m\bf k}}$.
Since we required all the transition functions be orientation-preserving,
the above relation shows that the determinant of the sewing matrix is uniform: $\det B_{CD}=\det(t^{AC})^{-1}\det B_{AB}\det t^{BD}=\det B_{AB}$.
Because $B=\pm i\sigma_y$ at time-reversal-invariant momenta (TRIM), such that $\det B=1$ at TRIM, the sewing matrix belongs to $SO(2)$ everywhere on the Brillouin zone.

The symmetry constraint on the Berry connection and curvature
\begin{align}
\tilde{\bf A}({\bf k})
&=-B^{T}({\bf k})\tilde{\bf A}({\bf -k})B({\bf k}) -B^{T}({\bf k})\nabla_{\bf k}B({\bf k}),\notag\\
\tilde{\bf F}({\bf k})
&=B^{T}({\bf k})\tilde{\bf F}({\bf -k})B({\bf k}).
\end{align}
reduce to
\begin{align}
\tilde{\bf A}({\bf k})+\tilde{\bf A}(-{\bf k})
&=
\begin{pmatrix}
0&-\nabla_{\bf k}\phi({\bf k})\\
\nabla_{\bf k}\phi({\bf k})&0
\end{pmatrix},\notag\\
\tilde{\bf F}({\bf k})
&=\tilde{\bf F}({\bf -k}),
\end{align}
where
\begin{align}
B({\bf k})
=
\begin{pmatrix}
\cos\phi({\bf k})&\sin \phi({\bf k})\\
-\sin \phi({\bf k})&\cos\phi({\bf k})
\end{pmatrix}.
\end{align}

Because the Fu-Kane-Mele invariant $\Delta$ is defined by the change of a 1D quantity, the time reversal polarization $P_T$, let us first investigate the 1D topological invariant.
Consider a time-reversal-invariant 1D subBrillouin zone, which includes two TRIM $\Gamma_1$ and $\Gamma_2$.
We can take a real smooth gauge there because the first Stiefel-Whitney class is trivial, i.e., the total Berry phase is trivial in complex smooth gauges as explained above.
On the time-reversal-invariant 1D Brillouin zone, we observe from symmetry conditions that
\begin{align}
\oint d{\bf k}\cdot \tilde{\bf A}_{12}({\bf k})
&=\int^{\Gamma_2}_{\Gamma_1} d{\bf k}\cdot (\tilde{\bf A}_{12}({\bf k}))+ \tilde{\bf A}_{12}(-{\bf k})))\notag\\
&=-\int^{\Gamma_2}_{\Gamma_1}d{\bf k}\cdot \nabla_{\bf k}\phi({\bf k})\notag\\
&=i\log\frac{\Pf B({\Gamma_2})}{\Pf B({\Gamma_1})}\text{ mod }2\pi\notag\\
&=2\pi P_T\text{ mod }2\pi,
\end{align}
where we used the definition of the time-reversal polarization in the last step~\cite{fu2006time}.
This integral is defined only modulo $2\pi$ because a gauge transformation can change its value by $2\pi$ times an integer
\footnote{Accordingly, $P_T$ here is gauge invariant modulo 1 while it depends on a gauge without $C_{2z}$ symmetry.
As $P_T$ is quantized and gauge invariant in the presence of $C_{2z}$, it serves as a topological mirror invariant~\cite{lau2016topological}.
Notice that $C_{2z}$ acts like a mirror on the time-reversal-invariant 1D subBrillouin zone.}.

Now we return to the original 2D Brillouin zone.
Let us take a real gauge where the occupied states are smooth over the region including the half Brillouin zone $0\le k_x\le \pi$ .
The Fu-Kane-Mele invariant $\Delta$ is defined as the time-reversal polarization pump from $k_x=0$ to $k_x=\pi$, i.e., $\Delta= P_T(\pi)-P_T(0)$, so
\begin{align}
\Delta
&=\frac{1}{2\pi}\left(i\log\frac{\Pf B(\pi,\pi)}{\Pf B(\pi,0)}-i\log\frac{\Pf B(0,\pi)}{\Pf B(0,0)}\right)\notag\\
&=\frac{1}{2\pi}\oint dk_y \tilde{A}_{12,y}(\pi,k_y)-\frac{1}{2\pi}\oint dk_y \tilde{A}_{12,y}(0,k_y)\notag\\
&=\frac{1}{2\pi}\int^{\pi}_0 dk_x\int^{\pi}_{-\pi}dk_y \tilde{F}_{12,z}(k_x,k_y)\notag\\
&=\frac{1}{4\pi}\int^{\pi}_{-\pi} dk_x\int^{\pi}_{-\pi}dk_y \tilde{F}_{12,z}(k_x,k_y)\notag\\
&=\frac{1}{2}e_2=\frac{1}{2}w_2\text{ mod }1,
\end{align}
where we used $\tilde{\bf F}(-{\bf k})=\tilde{\bf F}({\bf k})$ in the fourth line.
This shows the equivalence of the (two times) Fu-Kane-Mele invariant $\Delta$ and the second Stiefel-Whitney class $w_2$.

\section{Protection and characterization of the second-order topology}
\label{sec.SW-SOTI}

Recently, Wang {\it et al}. have proposed in~\cite{wang2018higher} that the anomalous corner charges are induced by the nontrivial second Stiefel-Whitney class.
Here review the idea in~\cite{wang2018higher} and discuss the effect of mirror and chiral symmetries.
Then, we establish the relation between the second Stiefel Whitney class and the nested Wilson loop~\cite{benalcazar2017quantized,benalcazar2017electric} employed in~\cite{wang2018higher} to capture the existence of the anomalous corner charges.

\subsection{Mirror and chiral symmetries}

For concrete discussion on the role of mirror and chiral symmetries, let us derive the presence of corner charges in Stiefel-Whitney insulators by reproducing the results in~\cite{wang2018higher}.

Consider a disk geometry with the radius $R$ shown in Fig.~\ref{fig.corner}(b) for simplicity.
For the sake of studying anomalous surface spectrum, it is enough to consider the low-energy effective Hamiltonian of the $I_{ST}$-symmetric doubled Chern insulator given by
\begin{align}
H_0({\bf r})
=-\Gamma_1i\d_x-\Gamma_2i\d_{y}+M({\bf r})\Gamma_3,
\end{align}
where $\Gamma_1=\tau_x$, $\Gamma_2=\tau_y\sigma_y$, $\Gamma_3=\tau_z$, and $I_{ST}=K$.
We take $M>0$ in the insulator and $M<0$ in the surrounding environment.
In polar coordinates,
\begin{align}
H_{0}(r,\theta)
=-\Gamma_1(\theta)i\d_r-\Gamma_2(\theta)r^{-1}i\d_{\theta}+M(r)\Gamma_3,
\end{align}
where $\Gamma_1(\theta)
=\cos\theta \Gamma_1+\sin\theta \Gamma_2$ and $\Gamma_2(\theta)
=-\sin\theta\Gamma_1+\cos\theta\Gamma_2$.
The surface degrees of freedom are given by projecting the matrices under
$P(\theta)=\frac{1}{2}(1+i\Gamma_1(\theta)\Gamma_3)$, and $H_0^{\rm edge}\equiv -(P\Gamma_2P)R^{-1}i\d_{\theta}|_{r=R}$ describes the two oppositely-propagating chiral edge states~\cite{khalaf2018higher,geier2018second,alicea2012new}.

Next, we consider perturbations to the Hamiltonian.
The terms that serve as edge mass are
\begin{align}
H_m(r,\theta)
&=m_{4}(r,\theta)\Gamma_{4}+m_{5}(r,\theta)\Gamma_{5}\notag\\
&+m_{24}(r,\theta)\Gamma_{24}(\theta)+m_{25}(r,\theta)\Gamma_{25}(\theta),
\end{align}
where $\Gamma_4
=\tau_y\sigma_x$,
$\Gamma_5
=\tau_y\sigma_z$,
$\Gamma_{2,j=4,5}(\theta)=-i\Gamma_2(\theta)\Gamma_j$.
One can easily see that all the matrices $\Gamma_4$, $\Gamma_5$, $\Gamma_{24}(\theta)$, and $\Gamma_{25}(\theta)$ anti-commute with $\Gamma_{2}(\theta)$ and commute with $P(\theta)$.
$I_{ST}$ symmetry $H_m^*(r,\theta)=H_m(r,\theta+\pi)$ imposes a constraint
\begin{align}
m_{I}(\theta+\pi)=-m_{I}(\theta)
\end{align}
for all $m_I=m_4,m_5,m_{24}$, and $m_{25}$.

The edge Hamiltonian is given by representing $P\Gamma_5P=-P\Gamma_{24}(\theta)P$, $P\Gamma_4P=P\Gamma_{25}(\theta)P$, and $P\Gamma_2P$ as $\tilde{\sigma}_x$, $\tilde{\sigma}$, and $\tilde{\sigma}_z$, respectively:
\begin{align}
H^{\rm edge}(\theta)
=\tilde{m}_{1}(\theta)\tilde{\sigma}_x+\tilde{m}_2(\theta)\tilde{\sigma}_y-R^{-1}i\d_{\theta}\tilde{\sigma}_z,
\end{align}
where $H^{\rm edge}(\theta)\equiv H^{\rm edge}_0+PH_mP|_{r=R}$, $\tilde{m}_1(\theta)=m_{4}(R,\theta)+m_{25}(R,\theta)$ and $\tilde{m}_2(\theta)=m_{5}(R,\theta)-m_{24}(R,\theta)$.
Each of mass terms $\tilde{m}_1$ and $\tilde{m}_2$ vanishes at least at even number of times due to the $I_{ST}$ symmetry constraint $\tilde{m}_{i=1,2}(\theta+\pi)=-m_{i}(\theta)$, but this does not require that edge band gap closes because the two mass terms do not simultaneously vanish in general.
However, if we require chiral symmetry $S(H+H_m)S^{-1}=-(H+H_m)$ where $S=\cos \eta \Gamma_4+\sin \eta \Gamma_5$, only one edge mass term remains.
Let us consider the case with $\eta=0$.
Then, $m_4=m_{25}=0$ such that $\tilde{m}_2=0$.
In this case, the only remaining mass $\tilde{m}_1$ should vanish at an even number of angle.
The domain kink therefore induces localized corner charges [See Fig.~\ref{fig.corner}(b)].

Let us consider mirror symmetry in addition to $I_{ST}$ and $S$ symmetries.
The mirror symmetry operator that gives
\begin{align}
M_yH_0(\theta)M^{-1}_y=H_0(-\theta)
\end{align}
is represented in the form of $M_y=\cos\chi \Gamma_{24}+\sin \chi \Gamma_{25}$.
When $\chi=0$, $M_y$ anti-commutes with $S=\Gamma_4$, and
\begin{align}
m_{5}(-\theta)
&=+m_5(\theta),\notag\\
m_{24}(-\theta)
&=+m_{24}(\theta),
\end{align}
and 
when $\chi=\pi$, $M_y$ commutes with $S=\Gamma_4$, and
\begin{align}
m_{5}(-\theta)
&=-m_5(\theta),\notag\\
m_{24}(-\theta)
&=-m_{24}(\theta).
\end{align}
Thus, mirror operator imposes that the edge mass flips sign at mirror-invariant corners only when it commutes with the chiral symmetry operator.
In addition, notice that $M_yI_{ST}:(x,y)\rightarrow (-x,y)$ acts like $M_x$ in the real space.
Combining the constraints given by $M_y$ and $I_{ST}$ one finds that
\begin{align}
m_{5}(\pi/2-\theta)
&=-m_5(\pi/2+\theta),\notag\\
m_{24}(\pi/2-\theta)
&=-m_{24}(\pi/2+\theta)
\end{align}
when $\chi=0$,
and 
\begin{align}
m_{5}(\pi/2-\theta)
&=+m_{5}(\pi/2+\theta),\notag\\
m_{24}(\pi/2-\theta)
&=+m_{24}(\pi/2+\theta)
\end{align}
when $\chi=\pi$.
Corner charges are localized at either $M_y$-invariant corners or at $M_x$-invariant corners but not at both.

\subsection{Tight-binding model}

\begin{figure}[t!]
\includegraphics[width=8.5cm]{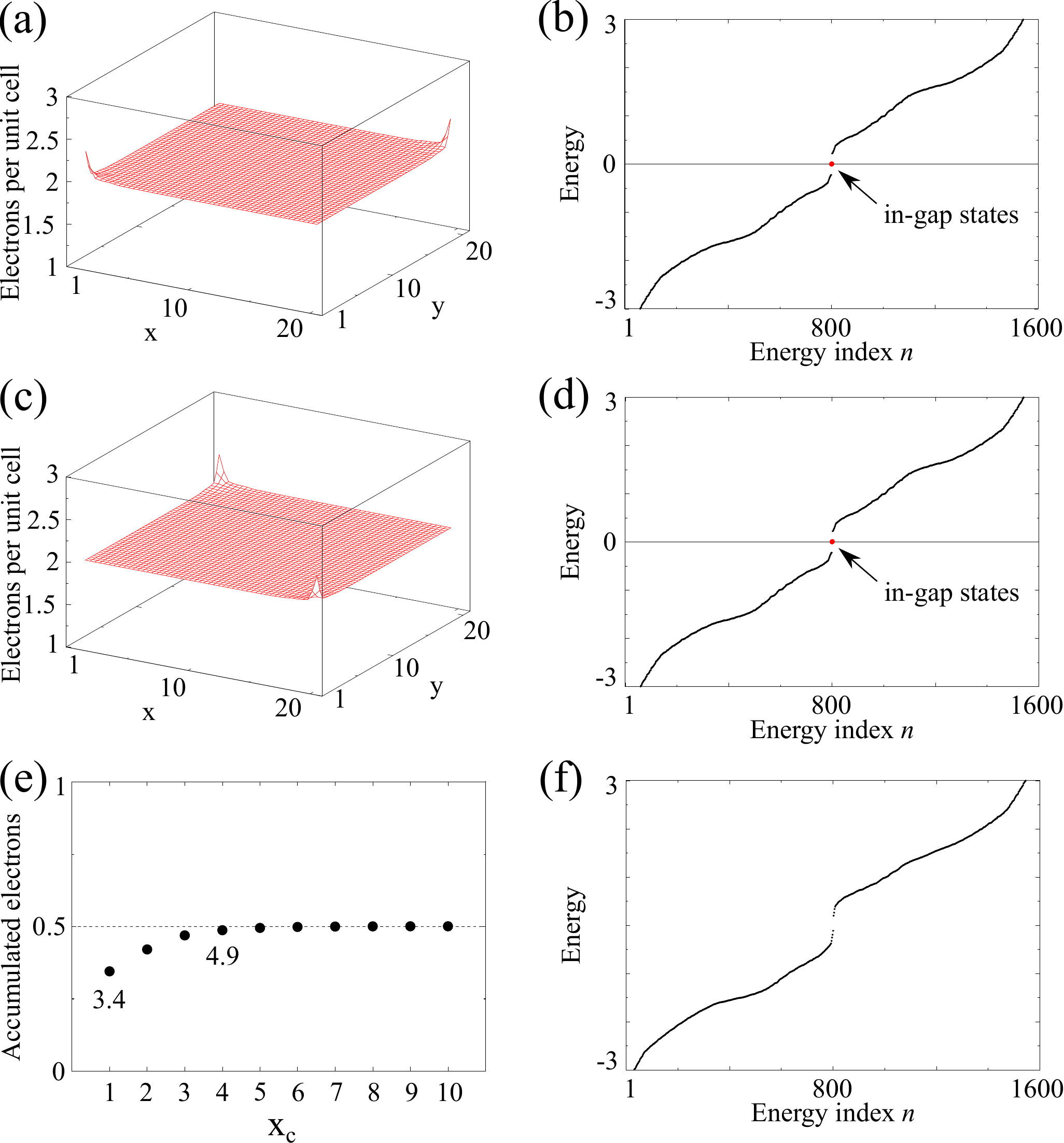}
\caption{
Corner charges in model Eq.~\eqref{eq.TB_Benjamin}.
Finite-size calculations are done by transforming the momentum space Hamiltonian into a square lattice tight-binding model, which has 20 by 20 unit cells.
(a,b) $M_{x+y}$ symmetric case. $m_1=m_3=0.4$ and $m_2=m_4=0.2$.
$\mu=0.01$ at $(x,y)=(1,1)$.
(c,d) $M_{x-y}$ symmetric case. $m_1=-m_3=0.4$ and $m_2=-m_4=0.2$. 
$\mu=0.01$ at $(x,y)=(20,0)$.
In (a) and (c), the Fermi level is assumed to be positive such that both corner charges are occupied.
(e) Accumulated electrons near the corner $(x,y)=(1,1)$ in (a).
It is calculated from $\sum^{x_c}_{x=1}\sum^{x_c}_{y=1}{\rho(x,y)-\braket{\rho(x,y)}}$, where $\rho(x,y)$ is the number of electrons in the unit cell at $(x,y)$, and $\braket{\rho(x,y)}=2$.
(f) In the absence of mirror symmetry. $m_1=0.1$, $m_3=0.4$ and $m_2=m_4=0.2$.
No in-gap states appear in this case.
}
\label{fig.corner_calculation}
\end{figure}

The model introduced in~\cite{wang2018higher} has the following form.
\begin{align}
\label{eq.TB_Benjamin}
H
&=\sin k_x\Gamma_1+\sin k_y\Gamma_2+(-1+\cos k_x+\cos ky)\Gamma_3\notag\\
&+m_1\Gamma_{14}+m_2\Gamma_{15}+m_{3}\Gamma_{24}+m_{4}\Gamma_{25},
\end{align}
where we defined three real Gamma matrices
\begin{gather}
\Gamma_1
=\tau_x,\quad
\Gamma_2
=\tau_y\sigma_y,\quad
\Gamma_3
=\tau_z,
\end{gather}
and two pure imaginary Gamma matrices
\begin{align}
\Gamma_4
=\tau_y\sigma_x,\quad
\Gamma_5
=\tau_y\sigma_z,
\end{align}
and the other generators of real matrices are then
\begin{align}
\Gamma_{14}
&=\tau_z\sigma_x,\quad
\Gamma_{15}
=\tau_z\sigma_z,\notag\\
\Gamma_{24}
&=-\sigma_z,\quad
\Gamma_{25}
=\sigma_x,\notag\\
\Gamma_{34}
&=-\tau_x\sigma_x,\quad
\Gamma_{35}
=-\tau_y\sigma_z.
\end{align}
The Hamiltonian is symmetric under
\begin{align}
P
=\Gamma_3,\quad
T
=\Gamma_3K.
\end{align}
$PT=K$ symmetry requires the Hamiltonian be real.

In~\cite{wang2018higher}, anomalous in-gap states were demonstrated with parameters, $m_1=0.3$, $m_{3}=0.4$, and $m_2=m_{4}=0.2$.
Let us note that this set of parameters are very close to the mirror and chiral symmetric parameters.
When $m_1=m_3$ and $m_2=m_4$, the Hamiltonian Eq.~\eqref{eq.TB_Benjamin} has chiral $S$ and two mirror $M_{x+y}:(x,y)\rightarrow (-y,-x)$ and $M_{x-y}: (x,y)\rightarrow (y,x)$ symmetries in addition to spatial inversion and time reversal symmetries.
To see this, let us rewrite the above Hamiltonian as
\begin{align}
H
&=\frac{1}{2}(\sin k_x+\sin k_y)(\Gamma_1+\Gamma_2)\notag\\
&+\frac{1}{2}(\sin k_x-\sin k_y)(\Gamma_1-\Gamma_2)\notag\\
&+(-3+\cos k_x+\cos ky)\Gamma_3\notag\\
&-i(\Gamma_1+\Gamma_2)(m_1\Gamma_4+m_2\Gamma_5).
\end{align}
In this form, one can see that it is symmetric under 
\begin{align}
M_{x+y}
&=\frac{i}{\sqrt{2m^2}}(\Gamma_1+\Gamma_2)(m_1\Gamma_4+m_2\Gamma_5),\notag\\
M_{x-y}
&=\frac{i}{\sqrt{2}m^2}(\Gamma_1-\Gamma_2)(m_1\Gamma_4-m_2\Gamma_5)(m_1\Gamma_4+m_2\Gamma_5),\notag\\
S
&=\frac{1}{\sqrt{m^2}}(m_1\Gamma_4+m_2\Gamma_5)
\end{align}
where $m^2=m_1^2+m_2^2$. $M_{x+y}^2=M_{x-y}^2=S^2=1$, and $M_{x+y}$, $M_{x-y}$, and $S$ all commutes with time reversal $T$.

According to the analysis in the previous subsection, corner charges are accumulated at the $M_{x-y}$-invariant corners when $m_1=m_3$ and $m_2=m_4$.
This is consistent with our calculations in Fig.~\ref{fig.corner_calculation}(a,b).
Also, if we choose parameters $m_1=-m_3$ and $m_2=-m_4$, corner charges are accumulated at $M_{x+y}$-invariant corners as shown in Fig.~\ref{fig.corner_calculation}(c,d) since then the role of $M_{x+y}$ and $M_{x-y}$ is interchanged.
The corner states carry half charges as shown in Fig.~\ref{fig.corner_calculation}(e).
Those in-gap states disappear when chiral symmetry is broken [See Fig.~\ref{fig.corner_calculation}(f)].

\subsection{Nested Wilson loop method}

\begin{figure}[t!]
\includegraphics[width=8.5cm]{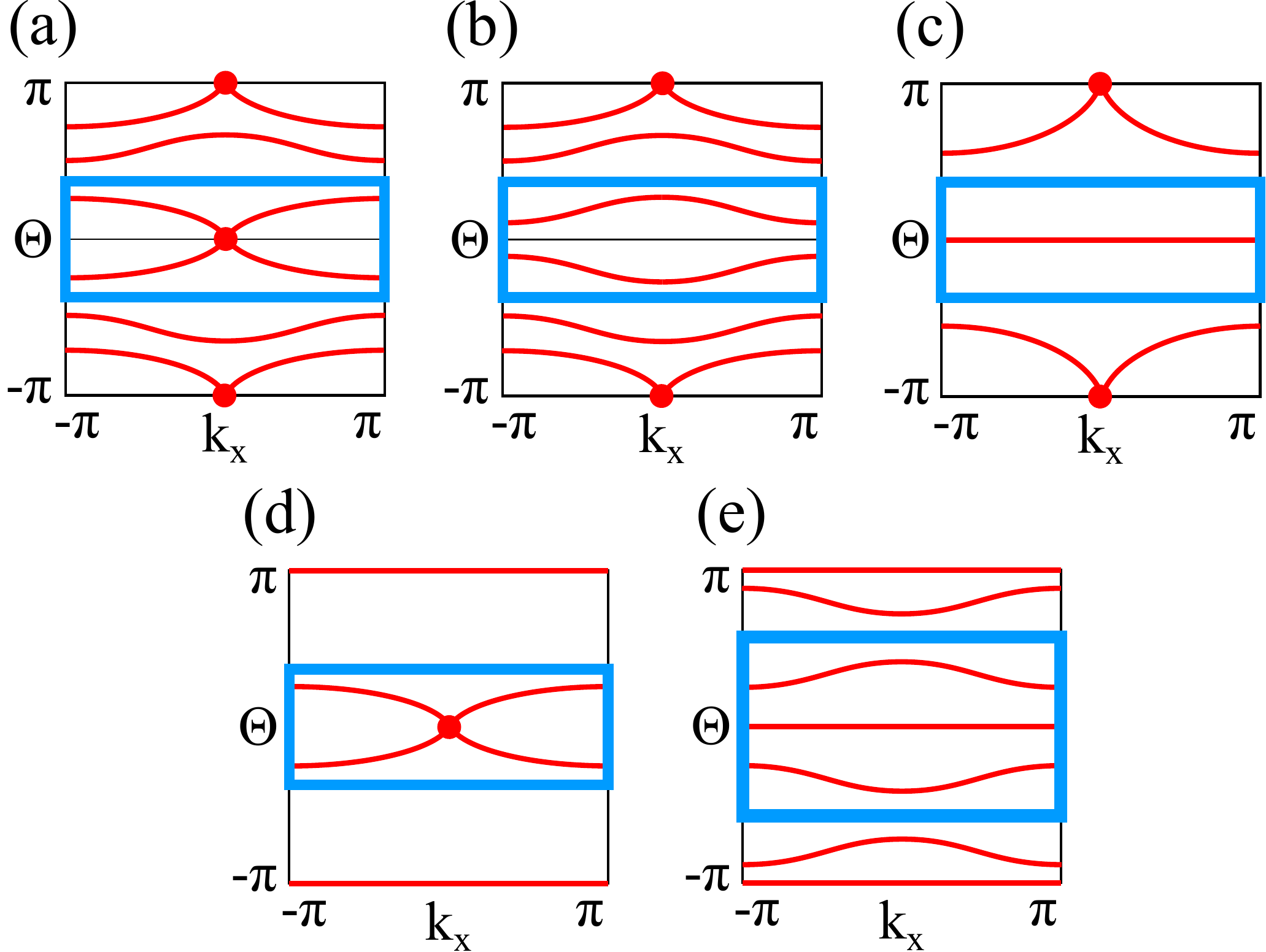}
\caption{
Wilson loop spectra with the nontrivial second Stiefel-Whitney class ($w_2=1$).
$\Theta(k_x)$ is the phase eigenvalue of the Wilson loop operator calculated along the $k_y$ direction at a fixed $k_x$. 
In each panel, a blue box indicates the block  ${\cal B}_1$ of Wilson bands centered at the $\Theta=0$ line whereas the rest of the Wilson bands form the other block ${\cal B}_2$ centered at the $\Theta=\pi$ line, each of which can be used to compute the nested Wilson loop $W_2({\cal B}_i)$ $(i=1,2)$ along the $k_x$ direction.
(a) $(\Phi_{x},\Phi_{y})=(0,0)$,
$\det W_2({\cal B}_1)=\det W_2({\cal B}_2)=-1$.
(b) $(\Phi_{x},\Phi_{y})=(\pi,0)$,
$\det W_2({\cal B}_1)=+1$, $\det W_2({\cal B}_2)=-1$.
(c) $(\Phi_{x},\Phi_{y})=(0,0)$ or $(\pi,0)$,
$\det W_2({\cal B}_1)=\pm 1$, $\det W_2({\cal B}_2)=-1$.
(d) $(\Phi_{x},\Phi_{y})=(0,\pi)$ or $(\pi,\pi)$,
$\det W_2({\cal B}_1)=-1$, $\det W_2({\cal B}_2)=\pm 1$.
(e) $(\Phi_{x},\Phi_{y})=(0,\pi)$ or $(\pi,\pi)$,
$\det W_2({\cal B}_1)=-1$, $\det W_2({\cal B}_2)=1$.
Here $\Phi_{x}$ and $\Phi_{y}$ are the Berry phases for the whole bands (usually the whole occupied bands) along the $k_x$ and $k_y$ directions, respectively.
In (a,b,c) where $\Phi_y=0$, $w_2$ can be determined by $\det W_2({\cal B}_2)$ whereas in (d,e) where $\Phi_y=\pi$, $w_2$ can be determined by $\det W_2({\cal B}_1)$. 
}
\label{fig.nested_Wilson}
\end{figure}

The nested Wilson loop method, originally proposed in ~\cite{benalcazar2017quantized,benalcazar2017electric}, was used in~\cite{wang2018higher} as a diagnostics for anomalous corner charges induced from the bulk topology.
Let us briefly recap the idea as follows.
First, one calculates the Wilson loop operator along the $k_y$ direction for a given momentum $k_x$. Here, $k_x$ and $k_y$ are arbitrary two independent momenta that parametrize the 2D Brillouin zone.
Then, its phase eigenvalues $\Theta$'s (so-called Wilson bands) are calculated as a function of $k_x$.
The spectrum is gapped in general except for possible crossings on the $\Theta=0$ or $\Theta=\pi$ lines, because only the crossings on the $\Theta=0$ and $\Theta=\pi$ lines are protected by $I_{ST}$ symmetry~\cite{ahn2018linking} [See Fig.~\ref{fig.nested_Wilson}].
Therefore, one can separate the Wilson bands into two groups ${\cal B}_1$ and ${\cal B}_2$ that are centered at $\Theta=0$ and $\Theta=\pi$, respectively, and separated by a gap inbetween.
The choice of ${\cal B}_1$ and ${\cal B}_2$ is not unique, and the number of bands in a group can vary depending on the choice. However, the topological characteristic of each group ${\cal B}_1$, ${\cal B}_2$ is independent of the choice.
Then we can pick a particular group ${\cal B}_i$ ($i=1,~2$), and calculate their determinant of the Wilson loop along $k_x$, i.e., the exponentiation of the Berry phase for ${\cal B}_i$ along $k_x$,
\begin{align}
\text{det}W_{2}({\cal B}_i)=\exp[i\Phi_{x}({\cal B}_i)].
\end{align}
It was suggested that this determinant of the ``nested Wilson loop" is $-1$ ($+1$) when an anomalous corner charge is (is not) present.
Let us note that this method can be used to the cases when the number of bands is bigger than two because the Wilson loop spectrum is not gapped in general for two bands. 
See Fig.~\ref{fig.TB}(d) in the main text as an example of a gapless Wilson loop spectrum for two bands.

Now we show how the nested Wilson loop is related to the second Stiefel-Whitney class, which is responsible for the appearance of the anomalous corner charges.
More specifically, let us clarify that i) which block of Wilson bands (${\cal B}_1$, or ${\cal B}_2$, or both) should be chosen to determine the second Stiefel-Whitney class $w_2$, and ii) whether it is possible to determine $w_2$ directly from the pattern of the Wilson loop spectrum without additional computation of the nested Wilson loop.
Notice that the decomposing the Wilson bands into two blocks ${\cal B}_1$ and ${\cal B}_2$ corresponds to decomposing the transition functions into two blocks, because the Wilson loop operator is equivalent to the transition functions in a parallel-transport gauge~\cite{ahn2018linking}.
Then, according to the Whitney sum formula~\cite{ahn2018linking,hatcher2003vector}, the total Stiefel-Whitney class $w_2({\cal B}_1\oplus {\cal B}_2)$ can be determined by the topological invariants of each block as
\begin{align}
&w_2({\cal B}_1\oplus {\cal B}_2)
=w_2({\cal B}_1)+w_2({\cal B}_2)\notag\\
&+\frac{1}{\pi^2}\left(\Phi_{x}({\cal B}_1)\Phi_y({\cal B}_2)+\Phi_{x}({\cal B}_2)\Phi_y({\cal B}_1)\right).
\end{align}

First, let us choose ${\cal B}_1$ to determine $\det W_2({\cal B}_1)=\exp[i\Phi_x({\cal B}_1)]$.
Notice that $\Phi_y({\cal B}_1)=0$ since ${\cal B}_1$ is centered at $\Theta=0$, and $w_2({\cal B}_1)=0$ since it is given by the number of the Wilson band crossings at the $\Theta=\pi$ line~\cite{ahn2018linking}.
Then we have
\begin{align}
&w_2({\cal B}_1\oplus {\cal B}_2)
=w_2({\cal B}_2)-\frac{i}{\pi^2}\log\left[\det W_2({\cal B}_1)\right]\Phi_y({\cal B}_2).
\end{align}

In fact, $w_2({\cal B}_2)$ and $\det W_2({\cal B}_1)$ can be further related to each other in some cases. 
In order to investigate all possible patterns of Wilson loop spectra with $w_2({\cal B}_1\oplus {\cal B}_2)=1$, as shown in Fig.~\ref{fig.nested_Wilson}, let us recall that the second Stiefel-Whitney class can be determined by the Wilson loop spectrum as follows~\cite{ahn2018linking}: 

i) when $\Phi_y=0$, it is given by the number of crossing points on the $\Theta=\pi$ line [Fig.~\ref{fig.nested_Wilson}(a,b,c)]; 

ii) when $\Phi_y=\pi$, it is given by the number of crossing points on the $\Theta=0$ line  if the number of bands is odd [Fig.~\ref{fig.nested_Wilson}(d)] whereas it is undetermined by the spectrum if the number of bands is even [Fig.~\ref{fig.nested_Wilson}(e)]; 

iii) furthermore, the Berry phase along $k_x$ direction $\Phi_x$ can be determined by the parity of the total number of crossing points on both the $\Theta=0$ and $\Theta=\pi$ lines. For instance, $\Phi_x=0$ in Fig.~\ref{fig.nested_Wilson}(a) ($\Phi_x=\pi$ in Fig.~\ref{fig.nested_Wilson}(b)) because there are even (odd) crossing points in total; 

iv) however, the Berry phase $\Phi_x$ is indeterminate when there are flat Wilson bands at $\Theta=0$ or $\pi$ as in Fig.~\ref{fig.nested_Wilson} (c),(d),(e).

Basically the same rule can be applied to a subset of Wilson bands, ${\cal B}_1$ or ${\cal B}_2$. Keeping the above rules in mind, let us consider the following two cases:

(1) When $\Phi_y({\cal B}_2)=0$, corresponding to the cases shown in Fig.~\ref{fig.nested_Wilson}(a,b,c), we also have $w_2({\cal B}_2)=1$ because of the single crossing point on the $\Theta=\pi$ line. Then $w_2({\cal B}_1\oplus {\cal B}_2)=1$ does not depend on $\det W_2({\cal B}_1)$.
Let us determine $\det W_2({\cal B}_1)$ by inspecting the evolution pattern of the Wilson bands in ${\cal B}_1$ and compare it with $w_2({\cal B}_1\oplus {\cal B}_2)=1$. 
Notice that $\det W_2({\cal B}_1)=-1$ and $+1$ in Fig.~\ref{fig.nested_Wilson} (a) and Fig.~\ref{fig.nested_Wilson} (b), respectively, since it can be determined by the number of the crossings on the $\Theta=0$ line, that is, $\det W_2=1$ ($-1$) when the number is even (odd)~\cite{ahn2018linking}. However, in the case shown in Fig.~\ref{fig.nested_Wilson} (c), $\det W_2({\cal B}_1)$ cannot be determined simply by looking at the shape of the Wilson band in ${\cal B}_1$. When there is a flat Wilson band on the $\Theta=0$ line, $\det W_2({\cal B}_1)$ should be determined directly via numerical computation.

Therefore, we find that although $w_2({\cal B}_1\oplus {\cal B}_2)=1$ is fixed for all the cases shown in Fig.~\ref{fig.nested_Wilson}(a,b,c), $\det W_2({\cal B}_1)$ varies depending on the shape of the Wilson bands in ${\cal B}_1$. So one cannot establish any relationship between $w_2({\cal B}_1\oplus {\cal B}_2)$ and $\det W_2({\cal B}_1)$. However, by using the other block ${\cal B}_2$, one can see that $\det W_2({\cal B}_2)=-1$ in all the three cases shown in Fig.~\ref{fig.nested_Wilson}(a,b,c). Thus we conclude that when $\Phi_y({\cal B}_1\oplus {\cal B}_2)=0$, $w_2({\cal B}_1\oplus {\cal B}_2)$ can be determined by $\det W_2({\cal B}_2)$.
This is basically because both are given by the parity of the crossing points at $\Theta=\pi$ in the non-nested Wilson loop spectrum. Namely, $w_2({\cal B}_1\oplus {\cal B}_2)=0~(1)$ indicates $\det W_2({\cal B}_2)=1~(-1)$ whereas $\det W_2({\cal B}_1)$ is not a meaningful quantity.

(2) When $\Phi_y({\cal B}_2)=\pi$, corresponding to the cases shown in Fig.~\ref{fig.nested_Wilson}(d,e),  $w_2({\cal B}_2)=0$ as the Wilson bands in ${\cal B}_2$ do not cross, so $w_2({\cal B}_1\oplus {\cal B}_2)=-\frac{i}{\pi}\log(\det W_2)$.
Thus $\det W_2({\cal B}_1)=-1$ gives $w_2({\cal B}_1\oplus {\cal B}_2)=1$ in these cases.
In Fig.~\ref{fig.nested_Wilson} (d), $\det W_2({\cal B}_1)=-1$ because of the odd number of Wilson band crossings on the $\Theta=0$ line.
On the other hand, in Fig.~\ref{fig.nested_Wilson} (e), $\det W_2({\cal B}_1)$ cannot be determined by the shape of the Wilson bands, so it should be directly calculated numerically. Therefore, we find that when $\Phi_y({\cal B}_1\oplus {\cal B}_2)=\pi$, $w_2({\cal B}_1\oplus {\cal B}_2)$ can be determined by $\det W_2({\cal B}_1)$, that is, $w_2({\cal B}_1\oplus {\cal B}_2)=0~(1)$ indicates $\det W_2({\cal B}_1)=1~(-1)$. It is straightforward to show that $\det W_2({\cal B}_2)$ is an irrelevant quantity in this case.

In conclusion, the determinant of the nested Wilson loop is equivalent to second Stiefel-Whitney class in all cases shown in Fig.~\ref{fig.nested_Wilson}. However, depending on the total Berry phase $\Phi_{y}({\cal B}_1\oplus {\cal B}_2)$, a different block of Wilson bands should be used to determine $w_2({\cal B}_1\oplus {\cal B}_2)$. Specifically, when $\Phi_{y}({\cal B}_1\oplus {\cal B}_2)$ is 0 ($\pi$), ${\cal B}_2$ (${\cal B}_1$) should be used to determine $\det W_2$ from the evolution pattern of the Wilson bands in the block. Let us note that, in the case shown in Fig.~\ref{fig.nested_Wilson} (e), $w_2$ cannot be determined simply by the shape of the (non-nested) Wilson loop spectrum, so the nested Wilson loop should be numerically calculated to get $w_2$.
In practice, however, there is an alternative way. We can avoid the case (e) by calculating the Wilson loop spectrum along a different direction (along $k_x$ or along $k_x+k_y$). Then $w_2$ can be determined directly by the pattern of the Wilson loop spectrum without additional numerical calculations of the nested Wilson loop.

\section{Fragility of $e_2=1$ phase in the presence of additional symmetries}
\label{sec.fragility_symmetries}

As we have discussed in Sec.~\ref{ssec.additional_symmetries}, the presence of $C_{3z}$ can fix the positions of the Dirac points. 
In such cases, it is no longer possible to prove the fragility of the $e_2=1$ phase by pair annihilating the Dirac points.
Instead, we should solve the fragility equation $(e_2=1)\oplus X=Y$ where $X$ and $Y$ are the sets of Wannier-representable bands.
To do this, let us organize the symmetry content, the representation content, and $w_2$ of an atomic insulator into a vector with 7 components, 
\begin{widetext}
\begin{equation}
\left[ C_{3z} \Gamma=1, C_{3z} \Gamma=e^{\pm 2\pi i/3}, C_{3z} K=1, C_{3z} K=e^{\pm 2\pi i/3}, C_{2x} \Gamma=1, C_{2x} \Gamma=-1, w_2\right].
\end{equation}
\end{widetext}
The first six components of this vector represent the number of bands with the specified $C_{3z}$ or $C_{2x}$ eigenvalues at the specified high symmetry points, while the last component of the vector contains the value of $w_2$.
For example, in the case of the nearly flat bands of TBG, whose symmetry and topology are captured by the lower two bands of the Hamiltonian given in Eq.~\eqref{eq.TB}, we have
\begin{equation}
(e_2=1)_{\rm TBG}=[2,0,0,2,1,1,1].
\end{equation}
Let us denote an atomic insulator with orbital ${\cal O}$ at the Wyckoff position  ${\cal W}$ by $({\cal W},{\cal O})$. For the minimal atomic insulators, we have~\cite{po2018faithful}
\begin{align*}
(A,s)= [1,0,1,0,1,0,0]\\
(A,p_z)= [1,0,1,0,0,1,0]\\
(A,p_\pm)= [0,2,0,2,1,1,0]\\
(B,s)= [2,0,0,2,2,0,0]\\
(B,p_z)= [2,0,0,2,0,2,0]\\
(B,p_\pm)= [0,4,2,2,2,2,0]\\
(C,s)= [1,2,1,2,2,1,1]\\
(C,p_z)= [1,2,1,2,1,2,1]\\
\end{align*}
Using the fact that $w_2$ is a $Z_2$ quantity, one finds that the minimal solutions are~\cite{po2018faithful}
\begin{equation}
(e_2=1)_{\rm TBG}+(C,p_z)=(A,s)+(A,p_\pm)+(B,p_z)
\end{equation}
and
\begin{equation}
(e_2=1)_{\rm TBG}+(C,s)=(A,p_z)+(A,p_\pm)+(B,s).
\end{equation}
There is no solution to the fragility equation if $X$ consists of one or two bands.


%

\end{document}